\documentclass[prd,nofootinbib,preprint,superscriptaddress,twocolumn,10pt]{revtex4}
\pdfoutput=1
\usepackage{aas_macros}
\usepackage{newtxtext,newtxmath}
\usepackage{hyperref}
\usepackage{booktabs}
\usepackage{mathtools}
\usepackage{cleveref}
\usepackage{makecell}
\usepackage{subfigure}
\usepackage{bm}
\usepackage{ulem}
\usepackage{orcidlink}
\newcommand{\be}{\begin{equation}}
\newcommand{\ee}{\end{equation}}
\newcommand{\ba}{\begin{eqnarray}}
\newcommand{\ea}{\end{eqnarray}}
\newcommand{\bi}{\begin{itemize}}
\newcommand{\ei}{\end{itemize}}

\usepackage{color}

\newcommand{\reffig}{Fig.\,\ref}

\usepackage{xcolor}
\DeclareRobustCommand{\VAN}[3]{#2}
\let\VANthebibliography\thebibliography
\def\thebibliography{\DeclareRobustCommand{\VAN}[3]{##3}\VANthebibliography}
\usepackage{graphicx}
\usepackage{grffile}% Including figure files
\begin{document}
% \label{firstpage}
% \pagerange{\pageref{firstpage}--\pageref{lastpage}}
% \maketitle
\title{Meta-Calibration of the Cosmic Magnification Coefficient: Toward Unbiased Weak Lensing Reconstruction by Counting Galaxies}

\author{Jian Qin\orcidlink{0000-0003-0406-539X}}
\email{qinjian@sjtu.edu.cn}
\affiliation{Department of Astronomy, School of Physics and Astronomy, Shanghai Jiao Tong University, Shanghai, 200240,China}
\affiliation{Key Laboratory for Particle Astrophysics and Cosmology
(MOE)/Shanghai Key Laboratory for Particle Physics and Cosmology,China}

\author{Pengjie Zhang\orcidlink{0000-0003-2632-9915}}
\email{zhangpj@sjtu.edu.cn}
\affiliation{Department of Astronomy, School of Physics and Astronomy, Shanghai Jiao Tong University, Shanghai, 200240,China}
\affiliation{Tsung-Dao Lee Institute, Shanghai Jiao Tong University, Shanghai, 200240, China}
\affiliation{Key Laboratory for Particle Astrophysics and Cosmology
(MOE)/Shanghai Key Laboratory for Particle Physics and Cosmology,China}

\author{Zhu Chen}
\affiliation{Shanghai Key Lab for Astrophysics, Shanghai Normal University, Shanghai 200234, China}

\author{Liping Fu}
\affiliation{Shanghai Key Lab for Astrophysics, Shanghai Normal University, Shanghai 200234, China}
\affiliation{Center for Astronomy and Space Sciences, China Three Gorges University, Yichang 443000, People’s Republic of China}

\author{Yu Yu\orcidlink{0000-0002-9359-7170}}
\email{yuyu22@sjtu.edu.cn}
\affiliation{Department of Astronomy, School of Physics and Astronomy, Shanghai Jiao Tong University, Shanghai, 200240,China}
\affiliation{Key Laboratory for Particle Astrophysics and Cosmology
(MOE)/Shanghai Key Laboratory for Particle Physics and Cosmology,China}

\author{Haojie Xu}
\affiliation{Shanghai Astronomical Observatory, Chinese Academy of Sciences, Shanghai 200030, China}
\affiliation{Department of Astronomy, School of Physics and Astronomy, Shanghai Jiao Tong University, Shanghai, 200240,China}
\affiliation{Key Laboratory for Particle Astrophysics and Cosmology
(MOE)/Shanghai Key Laboratory for Particle Physics and Cosmology,China}

\author{Ji Yao}
\affiliation{Shanghai Astronomical Observatory, Chinese Academy of Sciences, Shanghai 200030, China}
\affiliation{Department of Astronomy, School of Physics and Astronomy, Shanghai Jiao Tong University, Shanghai, 200240,China}
\affiliation{Key Laboratory for Particle Astrophysics and Cosmology
(MOE)/Shanghai Key Laboratory for Particle Physics and Cosmology,China}

\author{Yuan Shi}
\affiliation{Department of Astronomy, School of Physics and Astronomy, Shanghai Jiao Tong University, Shanghai, 200240,China}
\affiliation{Key Laboratory for Particle Astrophysics and Cosmology
(MOE)/Shanghai Key Laboratory for Particle Physics and Cosmology,China}

\author{Huanyuan Shan}
\affiliation{Shanghai Astronomical Observatory, Chinese Academy of Sciences, Shanghai 200030, China}

\begin{abstract}
Weak lensing alters galaxy sizes and fluxes, influencing the clustering patterns of galaxies through cosmic magnification. 
This effect enables the reconstruction of weak lensing convergence $\hat{\kappa}$ maps for DES and DECaLS by linearly combining galaxy overdensities across magnitude bins in the $g$, $r$, and $z$ photometry bands \citep{Qin+,Qin2+}. 
In this study, we enhance the lensing reconstruction method by addressing biases in the magnification coefficient estimation, which arise from incomplete consideration of selection effects, especially those induced by photometric redshift (photo-$z$) selection. 
Using a Random Forest-based photo-$z$ estimation for DECaLS and DES galaxies, we quantify the impact of photo-$z$ induced selection on magnification coefficient estimation. 
Our results show that neglecting photo-$z$ selection introduces significant biases in the magnification coefficient, leading to deviations in the reconstructed convergence map amplitude $A$, with values ranging from 0.4 to 3.5 depending on the survey, redshift, and magnitude cuts. 
By incorporating an improved magnification coefficient estimation that accounts for photo-$z$ selection, these biases are significantly reduced, with $A$ converging to $\sim 1$ as the magnitude cuts approach optimal values. 
This improvement is consistently observed across DES and DECaLS datasets and redshift bins, despite differences in survey strategies and depths. 
Our findings highlight the importance of addressing photo-$z$ induced selection to achieve unbiased weak lensing reconstructions and accurate cosmic magnification measurements.
\end{abstract}

\maketitle

%%%%%%%%%% BODY OF PAPER%%%%%
  \section{Introduction}

  Weak gravitational lensing is a powerful observational tool in cosmology that measures the subtle distortions of galaxy images (cosmic shear) or galaxy number densities (cosmic magnification).
  This phenomenon occurs due to the bending of light rays as they pass through gravitational field of foreground mass distributions. 
  Weak lensing studies have provided valuable insights into the large-scale structure of the universe, the nature of dark matter, and the properties of dark energy \citep{2001PhR...340..291B, 2015RPPh...78h6901K}.
  Cosmic shear are now contributing significantly to precision cosmology \citep[e.g.,][]{Hamana2020,2021A&A...645A.104A,2021A&A...645A.105G, 2022A&A...665A..56L,PhysRevD.105.023514,PhysRevD.105.023515, 2023arXiv230400702L}, under ongoing surveys such as the Dark Energy Survey \citep[DES,][]{DES2016}, the Kilo-Degree Survey \citep[KiDS,][]{KiDS2013}, the Hyper Suprime-Cam Subaru Strategic Program survey \citep[HSC-SSP,][]{HSC2018}, Vera C. Rubin Observatory's Legacy Survey of Space and Time \citep[LSST,][]{2009arXiv0912.0201L}, Euclid \citep{2011arXiv1110.3193L}, and the China Space Station Telescope (CSST) \citep{2019ApJ...883..203G,2024MNRAS.527.5206Y}. 
  Cosmic magnification, with improved observations, also shows great potential in weak lensing studies.

  One key advantage of cosmic magnification is that it directly affects the number density of galaxies, making it straightforward to measure compared to the subtle distortions in galaxy shapes. Therefore, a larger sample of galaxies can be used. For the same reason, cosmic magnification can be detected even at very high redshifts where shape measurements become challenging. Additionally, cosmic magnification is less sensitive to systematic errors related to the point spread function (PSF) and intrinsic alignments of galaxies, which are significant challenges in cosmic shear measurements. Combining the two measurements also allows for independent cross-checks of the results.

  In observations, cosmic magnification is typically detected through cross-correlations of two samples within the same sky area but at significantly different redshifts. Low-redshift lenses include luminous red galaxies (LRGs) and clusters \citep[e.g.,][]{pub.1059915677,2019MNRAS.484.1598B,2016MNRAS.457.3050C,2020MNRAS.495..428C}. High-redshift sources include quasars \citep{2005ApJ...633..589S,2012ApJ...749...56B}, Lyman break galaxies \citep{2012MNRAS.426.2489M,2017A&A...608A.141T}, and submillimetre galaxies \citep{2021A&A...656A..99B,Crespo_GonzalezNuevo_Bonavera_Cueli_Casas_Goitia_2022}. 
  However, these measurements of cosmic magnification are indirect, relying on cross-correlation.  Alternatively, it is also feasible to directly extract the magnification signal from multiple galaxy overdensity maps of different brightness \citep{2005PhRvL..95x1302Z,2011MNRAS.415.3485Y,ABS,YangXJ15,YangXJ17,Zhang18,2021RAA....21..247H,2024MNRAS.527.7547M}. The key aspect here to consider is the flux dependence characteristic of magnification bias. The main contamination to address is galaxy intrinsic clustering. Although galaxy bias is complex \citep[e.g.,][]{Bonoli09,Hamaus10,Baldauf10}, the primary component to eliminate is the deterministic bias. Ref.\,\cite{2021RAA....21..247H} introduced a modified internal linear combination (ILC) method that can remove the average galaxy bias in a model-independent way.

  In our recent studies \citep{Qin+,Qin2+}, we applied the reconstruction method to the Dark Energy Survey (DES) and the Dark Energy Camera Legacy Survey (DECaLS). 
  These efforts successfully suppressed galaxy intrinsic clustering to levels ranging from $\mathcal{O}(10^{-1})$ to $\mathcal{O}(10^{-2})$, achieving convergence-shear cross-correlation detections with signal-to-noise ratios of $10\sim 20$. 
  The magnification coefficient $g$ was estimated under the assumption of a flux-limited condition. 
  However, the survey selection functions often deviate from an ideal flux-limited condition, leading to biases in the reconstructed convergence map amplitudes.

  To address the biases in the magnification coefficient estimation, we need to comprehensively consider the real selection functions in the weak lensing analysis.  
  These selection functions are influenced by various factors, including imaging quality, environmental and morphological selection, as well as instrumental and redshift failures. 
  Previous studies \citep[e.g.,][]{Wietersheim-Kramsta_Joachimi_van,JElvinPoole2022DarkES,2024MNRAS.527.1760W} have explored these challenges in detail. 
  For instance, \citep{JElvinPoole2022DarkES} used imaging simulations to demonstrate that the flux-limited approximation is insufficient for accurate magnification coefficient estimation, resulting in biases ranging from $1.5$ to $3.5$, aligned with the biases observed in our DECaLS and DES cases.

  In this study, we focus on the photo-$z$ induced selection effects in DECaLS and DES galaxy samples. 
  Using a Random Forest algorithm, we estimate the photo-$z$ values of these galaxies, accounting for their sensitivity to galaxy flux and the influence by weak lensing magnification. 
  By analyzing this effect and quantifying its impact on the estimation of the magnification coefficient,
  we propose an improved magnification coefficient estimation method that accounts for the photo-$z$ induced selection.
  Our results demonstrate the effectiveness of this approach in achieving unbiased weak lensing reconstructions by counting galaxies.

  The paper is structured as follows. Section \ref{sec:methodology} introduces the RF photo-$z$ algorithm, explores the photo-$z$ induced selection effects, and evaluates their influence on the magnification coefficient estimation. Section \ref{sec:lensing_reconstruction} details the lensing reconstruction process, incorporating the improved magnification coefficient estimation, and validates the reconstruction methodology. Section \ref{sec:results} presents the results, highlighting the improvements in the reconstructed lensing convergence maps for DECaLS and DES galaxies. Finally, Section \ref{sec:summary} summarizes the findings and discusses potential future research directions.

  \begin{figure} 
  \centering
    \includegraphics[width=\columnwidth]{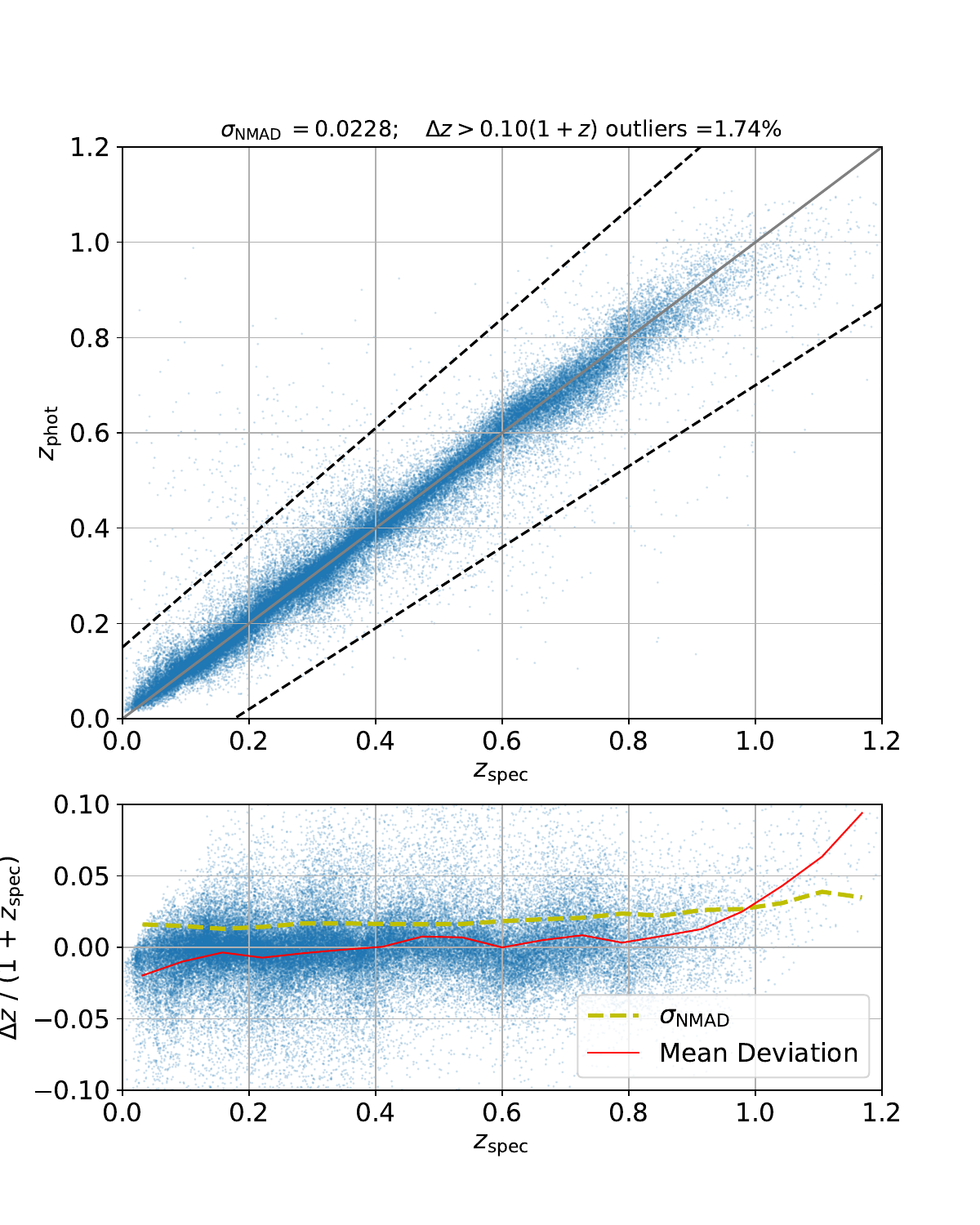}
  \caption{Photo-$z$ estimation performance.Top panel: Photo-$z$ vs. spec-$z$ for truth objects to show the photo-$z$ performance for $z_{\text{mag}} < 21.0$ objects in the DR9 South region. 
  Colors are used to indicate density where the points are densest. 
  Lower panel: photo-$z$ offset (in $\Delta z/(1 + z_{\text{spec}})$) vs. spec-$z$; the red solid line and the black dashed line are the mean deviation and $\sigma_{\text{NMAD}}$, respectively, of the photo-$z$ offset in bins of spec-$z$.}
  \label{fig:zspec_vs_zphot}
\end{figure}

\section{Methodology}\label{sec:methodology}

\subsection{Photometric redshift estimation}
Following \citet{zhou2021clustering}, we use the Random Forest (RF) algorithm to estimate the photo-$z$.
We use the same training set as in \citet{zhou2021clustering}, which consists of various redshift surveys overlapping with the DR9 south (DECaLS and DES) footprint, where they compile a redshift truth dataset using spectroscopic and many-band photometric redshifts from ten different surveys (see \citet{zhou2021clustering} for details).
We randomly select 90\% of the truth dataset for training and reserve the other 10\% for testing purposes.
We include the $r$-band magnitude as well as $g - r$, $r - z$, $z - W1$, and $W1 - W2$ colors as inputs.
To estimate the photo-$z$ error for each object, we perturb the photometry by adding to the observed flux in each band a random value from a Gaussian distribution whose standard deviation is set by the photometric error.
In \citet{zhou2021clustering}, morphological information (i.e., half-light radius, axis ratio, and shape probability) is also included as input, which was found to reduce the photo-$z$ scatter by 10\% to 20\%.
Our estimation of photo-$z$ is primarily utilized to investigate the photo-$z$ induced selection and understand how the RF predicted photo-$z$ of each galaxy is affected by weak lensing magnification. 
Specifically, we need to compare the photo-$z$ values of each galaxy with and without magnification.
However, the impact of magnification on the morphological information is not available in the current dataset. 
As a result, these features are not included in our photo-$z$ estimation, which slightly reduces the accuracy of the photo-$z$ predictions.
% This is also the reason why we do not use the ready-made photo-$z$ of \citet{zhou2021clustering} and take efforts to re-estimate the photo-$z$. 

\begin{figure} 
  \centering
    \includegraphics[width=1.05\columnwidth]{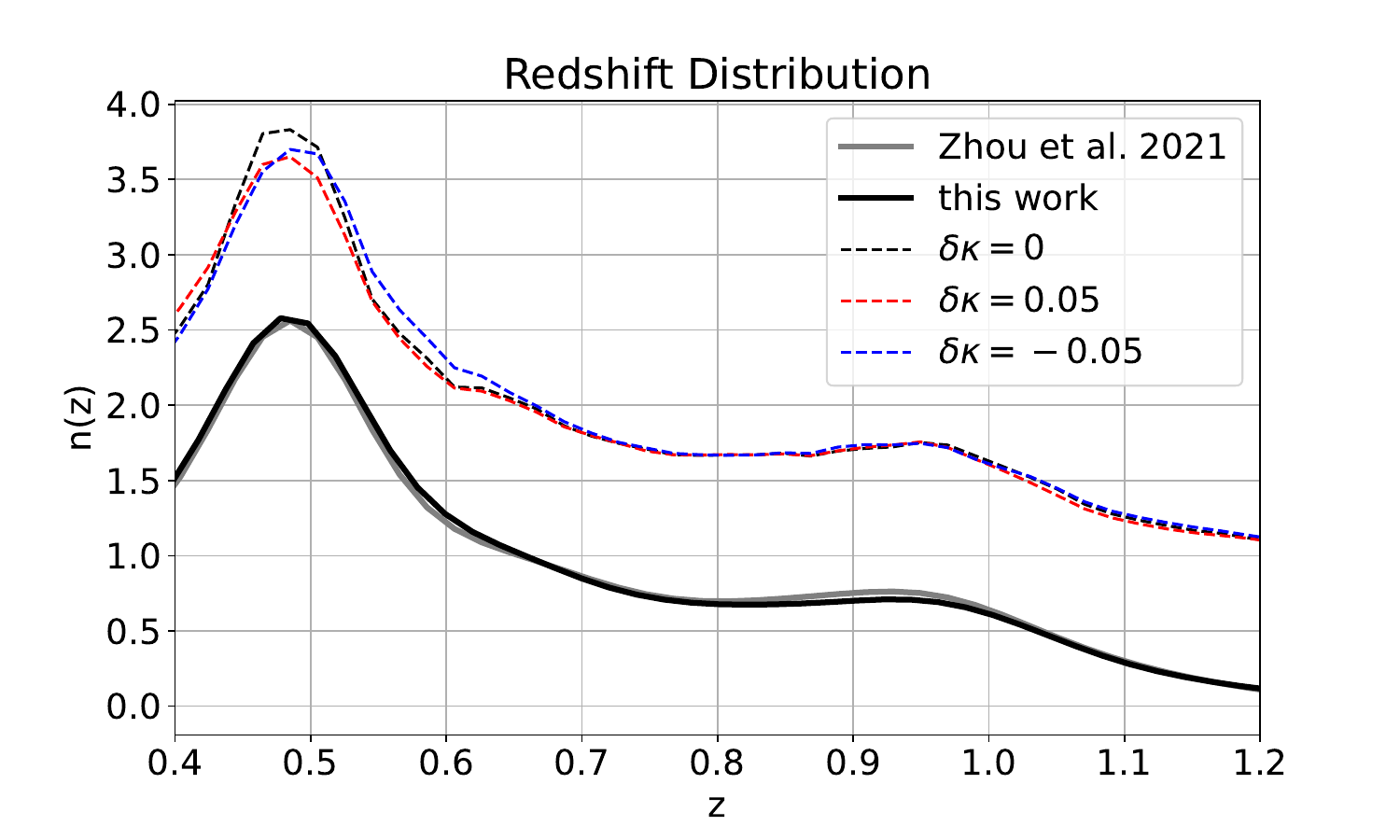}
    \caption{Normalized redshift distributions of DR9 South galaxies used for lensing reconstruction. 
    The black dashed line shows the raw distribution, while the black solid line represents the distribution convolved with photo-$z$ errors. 
    For comparison, the gray solid line shows the convolved distribution from \citet{zhou2021clustering}.
    The red and blue dashed lines illustrate the raw distributions of the same galaxies with a constant magnification ($\delta\kappa \sim \pm 0.05$) applied during photo-$z$ estimation.
    The distortion in the photo-$z$ distribution caused by magnification highlights the selection effect induced by photo-$z$.
    The dashed lines are shifted upward by +1 for clarity.
    Note that the distribution differences are significantly smaller for $|\delta\kappa| \leq 0.01$. 
    However, the magnification coefficient is determined not by the absolute amplitude of $|\delta\kappa|$, but by the rate of change of the selection function with respect to $\kappa$ (see Eq.\eqref{eq:g_definition}). To ensure the robustness of our analysis, we have verified that the numerical calculation of $g$ remains stable and converges for $|\delta\kappa| < 0.1$. For clarity and to better illustrate the selection effect, we adopt $\delta\kappa = \pm 0.05$ in this figure.
    }
  \label{fig:nz}
\end{figure}

\reffig{fig:zspec_vs_zphot} shows the relationship between photo-$z$ and spec-$z$ for $z_{\text{mag}} < 21.0$ objects in the truth catalog in the DR9 South region. 
The photo-$z$ scatter, quantified as $\sigma_{\text{NMAD}} = 1.48 \times \text{median}(|z_{\text{phot}} - z_{\text{spec}}|/(1 + z_{\text{spec}}))$, is 0.023.
The outlier rate, defined as the fraction of objects with $|z_{\text{phot}} - z_{\text{spec}}|/(1 + z_{\text{spec}}) > 0.1$, is 1.74\%.
These values are comparable to the photo-$z$ estimation results for $z_{\text{mag}} < 21.0$ galaxies of the DR8 South (DECaLS region) from \citet{zhou2021clustering}, which are 0.013 and 1.5\%, respectively.
\reffig{fig:nz} compares the redshift distributions for $g_{\text{mag}} < 24.0$ or $r_{\text{mag}} < 23.0$ or $z_{\text{mag}} < 22.5$ galaxies (corresponding to the faintest magnitude cuts for our lensing reconstruction, see Table \ref{tab:sub-sample}).
For consistency, we use our photo-$z$ for the redshift distribution estimation throughout this work.
We have also conducted tests and verified that using either redshift distribution yields negligible differences in the results.

\begin{figure*} 
  \includegraphics[width=1.75\columnwidth]{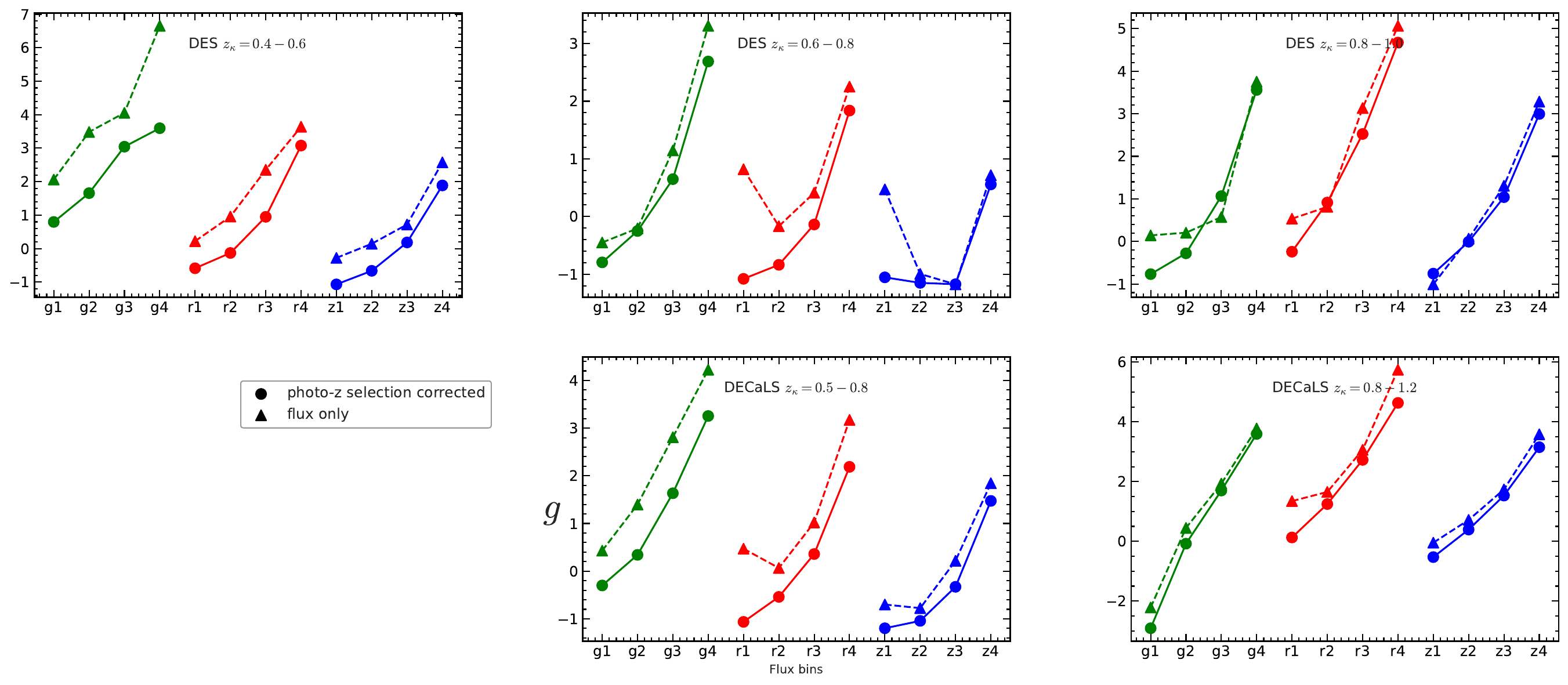}
  \caption{Magnification coefficient estimations for DES (top) and DECaLS (bottom) galaxies across flux bins in the $g$, $r$, and $z$ bands, and different photo-$z$ bins. 
  For each photo-$z$ bin, the faintest magnitude cuts are initially applied (see Table \ref{tab:sub-sample}). 
  Galaxies are then divided equally into four flux bins for each $g$, $r$, and $z$ band. 
  Solid lines represent the magnification coefficient estimation incorporating both photo-$z$ and flux-induced selection effects, while dashed lines show the estimation considering only flux-induced selection. 
  The discrepancy between the two estimations indicates the bias introduced by neglecting the photo-$z$ induced selection. 
  \label{fig:gs}}
\end{figure*}

  \begin{table}
    \centering
    \begin{tabular}{cccccccc}
    \hline
    \hline
  {} &  Photo-$z$ Bin &     Magnitude Cut in $g/r/z$       \\
  \midrule
  DECaLS  &    (0.5, 0.8) & 22.5/22.0/21.5 \\
   &    (0.8, 1.2) & 23.5/22.5/22.5  \\
  \hline
  &    (0.4, 0.6) & 22.5/22.0/21.5 \\
  DES  &    (0.6, 0.8) & 23.0/22.5/22.0  \\
  &    (0.8, 1.0) & 24.0/23.0/22.5 \\
  \hline
    \end{tabular}
    \caption{   Summary of the default magnitude cuts in $g$, $r$, and $z$ bands at different photo-$z$ bins for the DECaLS and DES galaxies.
    Similar to the magnitude cuts used in the baseline analysis by \citet{Qin+,Qin2+}, these cuts are determined by the peak position of the luminosity function.
    However, to address biases in the magnification coefficient estimation, we also investigate flux cuts shifted to brighter magnitudes, as shown in \reffig{fig:As}.
    }
    \label{tab:sub-sample}
  \end{table}

\begin{figure*} 
  \includegraphics[width=2.1\columnwidth]{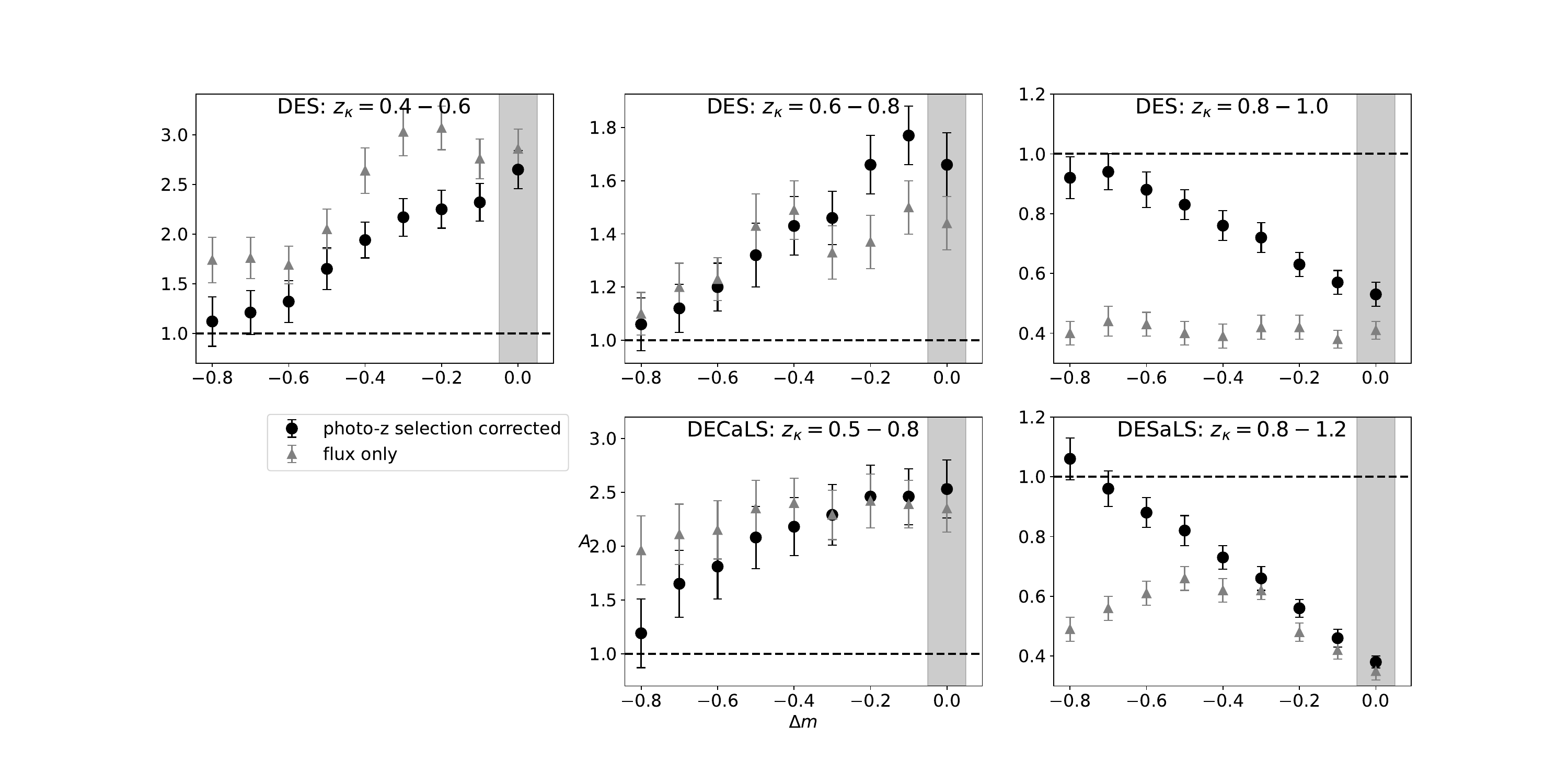}
  \caption{Constraints on $A$ of the reconstructed lensing convergence map as a function of the magnitude cuts $m_{g,r,z} + \Delta m$. 
  The top panels show results for DES at three redshift bins: $z_\kappa = 0.4-0.6$, $z_\kappa = 0.6-0.8$, and $z_\kappa = 0.8-1.0$. 
  The bottom panels show results for DECaLS at two redshift bins: $z_\kappa = 0.5-0.8$ and $z_\kappa = 0.8-1.2$.
  The black round data points represent constraints with the improved magnification coefficient estimation, while the grey triangle data points represent results with the flux-only estimation.
  The grey shaded regions correspond to $m_{g,r,z} + 0$, denoting the flux cuts determined by the peak position of the luminosity function.
  With the improved magnification coefficient estimation, $A$ converges nearly to one as the magnitude cuts shift to brighter values.
  In contrast, significant biases remain with the flux-only estimation, even with the brightest magnitude cuts ($\Delta m=-0.8$).
  \label{fig:As}}
\end{figure*}

\subsection{Photo-$z$ induced selection}

First, we recall how the magnification coefficient estimation is related to the survey selection function.
The overdensity due to convergence $\kappa$ at position $\hat{\mathbf{n}}$ on the sky can be written in terms of the observed number of galaxies, $N(\vec{F}, \hat{\mathbf{n}})$,
and the same quantity at $\kappa=0$ \citep[e.g.,][]{2010A&A...523A...1J,2009ApJ...695..652B}:
\begin{align}
\nonumber
&\delta^{\mathrm{mag}}(\hat{\mathbf{n}}, \kappa) = (1-2\kappa)\frac{N^{\mathrm{sel}}(\hat{\mathbf{n}}, \kappa)}{N^{\mathrm{sel}}(\hat{\mathbf{n}}, 0)} - 1 \ ,\\
\nonumber
&N^{\mathrm{sel}}(\hat{\mathbf{n}}, \kappa)=\int \mathrm{d} \vec{F} S(\vec{F}) N(\vec{F}, \hat{\mathbf{n}})\ .
\end{align}
Here, the superscript 'sel' indicates that a selection has been applied using thresholds on various observed properties of the galaxies, denote by $\vec{F}$. 
$S(\vec{F})$ is the sample selection function, which operates on galaxy properties $\vec{F}$. 
Magnification alters galaxy properties $\vec{F}$, such as flux and size.
Taylor expanding $S(\vec{F})$ around $\kappa=0$ and keeping the first-order terms yields \citep[e.g.,][]{2023MNRAS.523.3649E}:
\begin{align} \label{eq:g_definition}
&\delta^{\text {mag }}(\hat{\mathbf{n}}) = g\kappa(\hat{\mathbf{n}})\ , \\ \nonumber
&g = \frac{1}{N^{\mathrm{sel}}(\hat{\mathbf{n}}, 0)} \int \mathrm{d} \vec{F} \frac{\partial S}{\partial \kappa} N(\vec{F}, \hat{\mathbf{n}}) - 2\ .
\end{align}
Here, the '-2' term accounts for the change in the area element due to magnification, while the other term represents the response of the number of selected samples to a change in $\kappa$.

Real galaxy samples are selected based on a variety of properties, including flux, color, shape, position, and redshift. 
In this work, we focus on the selection based on photo-$z$. 
The selection operation on photo-$z$, $S(\vec{F}=\text{Photo-}z)$, typically involves binning galaxies within a certain photo-$z$ range. 
This selection will not affect the estimation of $g$ if cosmic magnification does not influence the photo-$z$ estimation, i.e., $\partial S(\vec{F}=\text{Photo-}z)/\partial \kappa = 0$  in Eq.\,\eqref{eq:g_definition}. 
However, photo-$z$ estimation is usually sensitive to galaxy flux, which is magnified by convergence $\kappa$, making $\partial S(\vec{F}=\text{Photo-}z)/\partial \kappa \neq 0$.

To illustrate this, we rerun the photo-$z$ estimation for the DR9 South galaxies with a constant magnification $\delta\kappa \sim \pm 0.05$ applied to each input galaxy. 
The comparison of the photo-$z$ distributions between the lensed and unlensed cases for the same input galaxies is shown in \reffig{fig:nz}. 
The differences in the distributions indicate the selection induced by photo-$z$, 
which contributes to the magnification coefficient estimation.

% The magnification coefficient $g$, under the flux-limited condition (i.e., the selection is only on the flux), is given by 
% \begin{equation}
%   g=2(\alpha - 1)\ ,\ \ \ \alpha=-\frac{d\ln{n(F)}}{d\ln{F}}-1\ .
%   \label{eq:prefactorg}
% \end{equation}

\subsection{Improved magnification coefficient estimation}

In our lensing reconstruction method \citep{Qin+,Qin2+}, galaxies are divided into flux and photo-$z$ bins. 
The faintest magnitude cuts in the $g$, $r$, and $z$ bands, along with the photo-$z$ bins, are summarized in Table \ref{tab:sub-sample}. 
For each photometric band, galaxies are equally divided into four flux bins. 
These binning induces two selections: $S(\vec{F}={\rm Flux}, \text{Photo-}z)$.

In \reffig{fig:gs}, we compare the magnification coefficient $g$ estimated considering $S(\vec{F}={\rm Flux}, \text{Photo-}z)$ against the estimation using the flux-only selection $S(\vec{F}={\rm Flux})$.
The latter approach was employed in our previous work \citep{Qin+,Qin2+}, but it is only unbiased under the flux-limited condition. 
The discrepancy between the two estimations of $g$ highlights the bias introduced by the flux-only estimation. 
This bias directly propagates to the amplitude of the reconstructed lensing convergence map.

\section{lensing reconstruction}\label{sec:lensing_reconstruction}

We enhance the lensing reconstruction for the DECaLS and DES galaxies using the improved magnification coefficient estimation. The details of the lensing reconstruction, data processing, imaging systematics mitigation, and convergence-shear cross-correlation analysis are described in \citep{Qin+,Qin2+}. Here, we summarize the key points.

The linear estimator for the convergence $\kappa$ is expressed as \citep{2021RAA....21..247H}:
\be\label{eq:linear combination}
\hat{\kappa}=\sum_{i} w_{i}\delta_{i}^{\rm L}\ , 
\ee 

The weights $w_i$ are determined by satisfying three conditions:
\begin{equation}
\label{eqn:wq}
  \text{Vanishing multiplicative error:} \sum_i w_i g_i=1\ ,
\end{equation}
\begin{equation}
\label{eqn:we}
\text{Eliminating intrinsic clustering:} \sum_i w_i =0\ ,
\end{equation}
\begin{equation}
\label{eqn:ws}
  \text{Minimizing shot noise:}\  N_{\rm shot}=\sum_{i,j}
w_{i}w_{j}\frac{\bar{n}_{ij}}{\bar{n}_{i}\bar{n}_{j}}\ .
\end{equation}

Here, $\delta_{i}^{\rm L}$ is the observed galaxy overdensity map in the $i$-th flux bin, and $\bar{n}_{ij}$ is the mean number density between flux bins $i$ and $j$. The magnification coefficient $g_i$ now accounts for both flux and photo-$z$ induced selection.

To characterize potential biases in the reconstructed $\hat{\kappa}$, we introduce parameters $A$ and $\epsilon$:
\be\label{eq:k+em}
\hat{\kappa}=A\kappa+\epsilon\delta_{\rm m}\ .
\ee
Here, $\epsilon$ represents the residual intrinsic clustering of galaxies, while $A$ accounts for any potential multiplicative error in the overall amplitude of $\hat{\kappa}$, which can arise from biases in $g$.

We then cross-correlate the reconstructed $\hat{\kappa}$ with cosmic shears:
\be\label{eq:xikg_decom}
{\xi}^{\kappa\gamma}_{j,\rm th}(\theta)=
A\xi^{\kappa\gamma}_j(\theta)+\epsilon\xi^{m\gamma}_j(\theta)\ ,
\ee
and compare the results with the theoretical models $\xi^{\kappa\gamma}$ and $\xi^{m\gamma}$ derived from the Planck 2018 cosmology. This comparison allows us to constrain the parameters $A$ and $\epsilon$.

\section{Results}\label{sec:results}
The constraints on $A$ for the DECaLS and DES $\kappa$ maps, using the flux-only estimation of the magnification coefficient, exhibit significant biases ranging from $1.2$ to $3.5$, depending on the $\kappa$ redshift, as reported in \citep{Qin+,Qin2+}. 
With the improved magnification coefficient estimation, the results, shown in \reffig{fig:As}, demonstrate a noticeable reduction in these biases. 
For reconstructions using the default magnitude cuts (determined by the peak of the luminosity function, as listed in Table \ref{tab:sub-sample}), the biases are moderately reduced. 
The improvement is particularly pronounced for $z_\kappa = 0.8-1.0$ in DES. 
However, despite these improvement, significant biases persist across all cases.

To better approximate the true selection function, we explore brighter magnitude cuts by shifting the default values $m_{g,r,z}$ to $m_{g,r,z} + \Delta m$, where $\Delta m$ ranges from $-0.8$ to $0$. 
\reffig{fig:As} shows the constraints on $A$ for these shifted flux cuts. 
With the improved magnification coefficient estimation, the biases in $A$ are significantly reduced as the magnitude cut becomes brighter. 
Notably, $A$ converges to $\sim 1$ as the magnitude cut approaches the brightest end ($\Delta m = -0.8$), consistent across DES, DECaLS, and all redshift bins. 
In contrast, the flux-only estimation retains significant biases, even at $\Delta m = -0.8$.

Tables \ref{tab:des} and \ref{tab:decals} present detailed constraints on $A$ and $\epsilon$ across various redshift bins and $\Delta m$ values, showcasing robust goodness-of-fit ($\chi^2_{\rm min}/{\rm d.o.f} \sim 1$) and effective suppression of intrinsic clustering ($\epsilon \sim 0$). We conclude that the improved magnification coefficient estimation leads toward an unbiased weak lensing reconstruction by counting galaxies. The results also indicate that simply shifting the magnitude cuts to brighter values is unable to approximate the true flux-limited condition.
Addressing the photo-$z$ induced selection is a critical step toward achieving an unbiased estimation.

For fainter magnitude cuts ($-0.7 \leq \Delta m \leq 0$), residual biases persist even with the improved magnification coefficient estimation. This suggests that photo-$z$ induced selection is not the only source of bias in these cases. Other factors, such as galaxy size, shape, imaging quality, and instrument or redshift failures, may contribute to additional selection effects. Addressing these issues requires a comprehensive analysis of all potential selection functions, encompassing the entire pipeline from initial imaging to the final galaxy catalog construction. We leave this detailed investigation for future work.

\begin{table*}
  \centering
  \begin{tabular}{lllllll}
  \toprule
  \hline\hline
  {} &                           $A$  &                      $\epsilon$ & $\chi^2_{\rm min}/{\rm d.o.f}$ & $\sqrt{\chi^2_{\rm null}-\chi^2_{\rm min}}$ & $A/\sigma_A$ & $z_\kappa$ \\
  \midrule
\hline
$\delta m = -0.8$ &  1.12$\pm$0.25(1.74$\pm$0.23) &   0.01$\pm$0.02(-0.03$\pm$0.02) &           1.0(1.2) &                                   8.6(10.9) &     4.5(7.6) \\
-0.7              &  1.21$\pm$0.22(1.76$\pm$0.21) &  -0.01$\pm$0.01(-0.05$\pm$0.01) &           0.9(1.0) &                                   9.2(11.1) &     5.5(8.4) \\
-0.6              &  1.32$\pm$0.21(1.69$\pm$0.19) &  -0.01$\pm$0.01(-0.05$\pm$0.01) &           0.8(0.9) &                                   9.4(11.3) &     6.3(8.9) \\
-0.5              &  1.65$\pm$0.21(2.05$\pm$0.20) &  -0.04$\pm$0.01(-0.07$\pm$0.01) &           0.9(1.0) &                                  10.1(12.0) &    7.9(10.2) \\
-0.4              &  1.94$\pm$0.18(2.64$\pm$0.23) &  -0.05$\pm$0.01(-0.09$\pm$0.01) &           1.1(1.3) &                                  13.1(13.3) &   10.8(11.5) & 0.4-0.6 \\
-0.3              &  2.17$\pm$0.19(3.03$\pm$0.24) &  -0.07$\pm$0.01(-0.11$\pm$0.02) &           1.1(1.3) &                                  13.6(14.3) &   11.4(12.6) \\
-0.2              &  2.25$\pm$0.19(3.07$\pm$0.22) &  -0.08$\pm$0.01(-0.12$\pm$0.01) &           1.4(1.7) &                                  13.4(15.3) &   11.8(13.9) \\
-0.1              &  2.32$\pm$0.19(2.76$\pm$0.20) &  -0.08$\pm$0.01(-0.11$\pm$0.01) &           1.3(1.8) &                                  14.3(15.7) &   12.2(13.8) \\
0.0               &  2.65$\pm$0.19(2.86$\pm$0.20) &  -0.10$\pm$0.01(-0.12$\pm$0.01) &           1.5(1.8) &                                  15.5(15.3) &   13.9(14.3) \\
\hline
$\delta m = -0.8$ &  1.06$\pm$0.10(1.10$\pm$0.08) &   0.02$\pm$0.02(-0.00$\pm$0.01) &           1.0(0.9) &                                  17.9(20.1) &   10.6(13.8) \\
-0.7              &  1.12$\pm$0.09(1.20$\pm$0.09) &   0.01$\pm$0.02(-0.02$\pm$0.01) &           1.1(0.8) &                                  19.0(20.0) &   12.4(13.3) \\
-0.6              &  1.20$\pm$0.09(1.23$\pm$0.08) &  -0.01$\pm$0.01(-0.03$\pm$0.01) &           1.0(0.9) &                                  20.0(20.6) &   13.3(15.4) \\
-0.5              &  1.32$\pm$0.12(1.43$\pm$0.12) &  -0.01$\pm$0.02(-0.04$\pm$0.02) &           1.0(0.8) &                                  16.4(16.7) &   11.0(11.9) \\
-0.4              &  1.43$\pm$0.11(1.49$\pm$0.11) &  -0.02$\pm$0.02(-0.05$\pm$0.02) &           0.8(0.7) &                                  18.8(17.8) &   13.0(13.6) & 0.6-0.8 \\
-0.3              &  1.46$\pm$0.10(1.33$\pm$0.10) &  -0.03$\pm$0.02(-0.05$\pm$0.01) &           0.7(0.5) &                                  18.7(17.4) &   14.6(13.3) \\
-0.2              &  1.66$\pm$0.11(1.37$\pm$0.10) &  -0.06$\pm$0.02(-0.05$\pm$0.01) &           1.0(0.6) &                                  18.9(18.4) &   15.1(13.7) \\
-0.1              &  1.77$\pm$0.11(1.50$\pm$0.10) &  -0.07$\pm$0.02(-0.06$\pm$0.02) &           1.1(0.8) &                                  19.2(18.9) &   16.1(15.0) \\
0.0               &  1.66$\pm$0.12(1.44$\pm$0.10) &  -0.06$\pm$0.02(-0.05$\pm$0.02) &           0.9(0.8) &                                  18.3(18.3) &   13.8(14.4) \\
\hline
$\delta m = -0.8$ &  0.92$\pm$0.07(0.40$\pm$0.04) &  -0.04$\pm$0.05(0.02$\pm$0.04) &           1.2(1.2) &                                  16.2(11.6) &   13.1(10.0) \\
-0.7              &  0.94$\pm$0.06(0.44$\pm$0.05) &  -0.04$\pm$0.05(0.01$\pm$0.04) &           1.1(1.2) &                                  18.5(12.0) &    15.7(8.8) \\
-0.6              &  0.88$\pm$0.06(0.43$\pm$0.04) &  -0.04$\pm$0.04(0.02$\pm$0.04) &           0.9(1.1) &                                  19.2(12.4) &   14.7(10.8) \\
-0.5              &  0.83$\pm$0.05(0.40$\pm$0.04) &  -0.02$\pm$0.04(0.03$\pm$0.03) &           0.8(1.2) &                                  19.6(12.6) &   16.6(10.0) \\
-0.4              &  0.76$\pm$0.05(0.39$\pm$0.04) &  -0.01$\pm$0.04(0.04$\pm$0.03) &           0.7(1.4) &                                  19.7(12.9) &    15.2(9.8) & 0.8-1.0 \\
-0.3              &  0.72$\pm$0.05(0.42$\pm$0.04) &  -0.02$\pm$0.04(0.03$\pm$0.03) &           1.0(1.5) &                                  19.2(14.6) &   14.4(10.5) \\
-0.2              &  0.63$\pm$0.04(0.42$\pm$0.04) &  -0.01$\pm$0.04(0.01$\pm$0.03) &           0.9(1.5) &                                  18.4(14.6) &   15.8(10.5) \\
-0.1              &  0.57$\pm$0.04(0.38$\pm$0.03) &   0.01$\pm$0.03(0.02$\pm$0.03) &           0.9(1.2) &                                  18.7(15.5) &   14.2(12.7) \\
0.0               &  0.53$\pm$0.04(0.41$\pm$0.03) &  -0.01$\pm$0.03(0.02$\pm$0.02) &           0.9(1.4) &                                  18.0(18.0) &   13.2(13.7) \\

  \bottomrule
  \end{tabular}
  \caption{Summary of the constraints on $A$ and $\epsilon$ for the DES convergence maps reconstructed at the redshift bins $z_\kappa = 0.4-0.6$, $z_\kappa = 0.6-0.8$, and $z_\kappa = 0.8-1.0$.
    Different flux cuts $m_{g,r,z}+\Delta m$ are considered.
    $\Delta m=0$ corresponds to the default flux cuts determined by the peak position of the luminosity function.
    The main table presents the results obtained using the improved magnification coefficient estimation, while the values in parentheses correspond to the results from the flux-only estimation.
    With the improved magnification coefficient estimation, $A$ converges to one as $\Delta m$ reaches $-0.8$.
    However, significant biases remain with the flux-only estimation, even with the brightest magnitude cuts ($\delta m=-0.8$).
    }
  \label{tab:des}
  \end{table*}

\begin{table*}
  \centering
  \begin{tabular}{lllllll}
  \toprule
  \hline\hline
  {} &                        $A$ &     $\epsilon$ & $\chi^2_{\rm min}/{\rm d.o.f}$ & $\sqrt{\chi^2_{\rm null}-\chi^2_{\rm min}}$ & $A/\sigma_A$ & $z_\kappa$ \\
  \midrule
\hline
$\delta m = -0.8$ &  1.19$\pm$0.32(1.96$\pm$0.32) &  0.11$\pm$0.03(0.09$\pm$0.03) &           0.8(0.8) &                                  12.2(15.2) &     3.7(6.1) \\
-0.7              &  1.65$\pm$0.31(2.11$\pm$0.28) &  0.11$\pm$0.03(0.06$\pm$0.03) &           1.1(0.9) &                                  15.7(17.0) &     5.3(7.5) \\
-0.6              &  1.81$\pm$0.30(2.15$\pm$0.27) &  0.10$\pm$0.03(0.05$\pm$0.02) &           1.0(0.9) &                                  17.9(18.7) &     6.0(8.0) \\
-0.5              &  2.08$\pm$0.29(2.35$\pm$0.26) &  0.07$\pm$0.03(0.04$\pm$0.02) &           0.8(0.9) &                                  17.9(19.6) &     7.2(9.0) \\
-0.4              &  2.18$\pm$0.27(2.40$\pm$0.23) &  0.06$\pm$0.03(0.03$\pm$0.02) &           0.9(1.0) &                                  18.1(20.5) &    8.1(10.4) & 0.5-0.8 \\
-0.3              &  2.29$\pm$0.28(2.29$\pm$0.23) &  0.05$\pm$0.03(0.03$\pm$0.02) &           0.8(1.1) &                                  18.0(20.3) &    8.2(10.0) \\
-0.2              &  2.46$\pm$0.29(2.42$\pm$0.25) &  0.03$\pm$0.03(0.01$\pm$0.02) &           0.7(1.2) &                                  16.9(18.7) &     8.5(9.7) \\
-0.1              &  2.46$\pm$0.26(2.39$\pm$0.22) &  0.03$\pm$0.03(0.01$\pm$0.02) &           0.7(1.4) &                                  18.8(21.4) &    9.5(10.9) \\
0.0               &  2.53$\pm$0.27(2.35$\pm$0.22) &  0.02$\pm$0.03(0.01$\pm$0.02) &           0.7(1.4) &                                  19.0(21.4) &    9.4(10.7) \\
\hline
$\delta m = -0.8$ &  1.06$\pm$0.07(0.49$\pm$0.04) &   -0.12$\pm$0.07(0.00$\pm$0.04) &  1.3(1.2) &  18.8(15.8) &  15.1(12.2) \\
-0.7              &  0.96$\pm$0.06(0.56$\pm$0.04) &  -0.04$\pm$0.07(-0.01$\pm$0.04) &  1.0(1.1) &  19.7(18.1) &  16.0(14.0) \\
-0.6              &  0.88$\pm$0.05(0.61$\pm$0.04) &  -0.07$\pm$0.05(-0.00$\pm$0.04) &  1.3(1.3) &  20.8(20.6) &  17.6(15.2) \\
-0.5              &  0.82$\pm$0.05(0.66$\pm$0.04) &   -0.02$\pm$0.05(0.01$\pm$0.04) &  1.3(1.2) &  21.0(21.7) &  16.4(16.5) \\
-0.4              &  0.73$\pm$0.04(0.62$\pm$0.04) &   -0.01$\pm$0.05(0.02$\pm$0.04) &  1.2(1.4) &  22.0(21.9) &  18.2(15.5) & 0.8-1.2 \\
-0.3              &  0.66$\pm$0.04(0.62$\pm$0.03) &  -0.04$\pm$0.04(-0.03$\pm$0.04) &  1.2(1.3) &  21.9(22.7) &  16.5(20.7) \\
-0.2              &  0.56$\pm$0.03(0.48$\pm$0.03) &   -0.06$\pm$0.04(0.02$\pm$0.03) &  1.3(1.4) &  20.3(20.1) &  18.7(16.0) \\
-0.1              &  0.46$\pm$0.03(0.42$\pm$0.03) &  -0.04$\pm$0.03(-0.01$\pm$0.03) &  1.7(1.9) &  19.0(17.5) &  15.3(14.0) \\
0.0               &  0.38$\pm$0.02(0.35$\pm$0.03) &   -0.04$\pm$0.03(0.01$\pm$0.04) &  3.2(3.4) &  18.4(15.8) &  19.0(11.7) \\
\bottomrule
  \end{tabular}
  \caption{Same as Table \ref{tab:des}, but for the DECaLS convergence maps reconstructed at  $z_\kappa = 0.5-0.8$ and $z_\kappa = 0.8-1.2$.  
  }
  \label{tab:decals}
  \end{table*}
  
  %put the figures Clkk and Wiener, two plots in one row  

  \section{Summary}\label{sec:summary}

  This study investigates the impact of photo-$z$ induced selection on the estimation of the magnification coefficient using a Random Forest-based algorithm. Since photo-$z$ estimation is sensitive to galaxy flux, which is influenced by cosmic magnification, it introduces an additional selection effect beyond the flux-limited condition. Neglecting this effect leads to biases in the magnification coefficient estimation and the reconstructed lensing convergence maps.

  By incorporating the photo-$z$ induced selection, we improved the magnification coefficient estimation, which significantly reduced the multiplicative biases in the reconstructed convergence maps for both DECaLS and DES. The improvement is particularly significant with brighter magnitude cuts, achieving nearly unbiased results when the cuts are shifted 0.8 magnitudes brighter than the luminosity function peak.

  This work primarily addresses photo-$z$ induced selection. 
  Future research should aim to account for the complete selection function, which can be derived from the entire pipeline of galaxy catalog construction.
  As a next step, we plan to investigate this using image simulations, such as the Balrog \citep[e.g.,][]{2022ApJS..258...15E} simulations developed for DES, as well as simulations currently under preparation for CSST.

Our method is not limited to the Random Forest-derived photo-$z$; instead, it represents a general approach applicable to any photo-$z$ algorithm. 
Incorporating more comprehensive photo-$z$ estimations, such as those including additional features like size and morphology, will achieve better photo-$z$ accuracy and lower outlier rate. 
These features introduce additional selection effects that must be accounted for in the estimation of magnification coefficient.

\section*{Acknowledgements}
This work is supported the National Key R\&D Program of China (2023YFA1607800, 2023YFA1607801, 2023YFA1607802, 2020YFC2201602), the China Manned Space Project (\#CMS-CSST-2021-A02), and the Fundamental Research Funds for the Central Universities.

\bibliography{mybib}

\begin{thebibliography}{47}
\expandafter\ifx\csname natexlab\endcsname\relax\def\natexlab#1{#1}\fi
\expandafter\ifx\csname bibnamefont\endcsname\relax
  \def\bibnamefont#1{#1}\fi
\expandafter\ifx\csname bibfnamefont\endcsname\relax
  \def\bibfnamefont#1{#1}\fi
\expandafter\ifx\csname citenamefont\endcsname\relax
  \def\citenamefont#1{#1}\fi
\expandafter\ifx\csname url\endcsname\relax
  \def\url#1{\texttt{#1}}\fi
\expandafter\ifx\csname urlprefix\endcsname\relax\def\urlprefix{URL }\fi
\providecommand{\bibinfo}[2]{#2}
\providecommand{\eprint}[2][]{\url{#2}}

\bibitem[{\citenamefont{{Qin} et~al.}(2023)\citenamefont{{Qin}, {Zhang}, {Xu},
  {Yu}, {Yao}, {Ma}, and {Shan}}}]{Qin+}
\bibinfo{author}{\bibfnamefont{J.}~\bibnamefont{{Qin}}},
  \bibinfo{author}{\bibfnamefont{P.}~\bibnamefont{{Zhang}}},
  \bibinfo{author}{\bibfnamefont{H.}~\bibnamefont{{Xu}}},
  \bibinfo{author}{\bibfnamefont{Y.}~\bibnamefont{{Yu}}},
  \bibinfo{author}{\bibfnamefont{J.}~\bibnamefont{{Yao}}},
  \bibinfo{author}{\bibfnamefont{R.}~\bibnamefont{{Ma}}}, \bibnamefont{and}
  \bibinfo{author}{\bibfnamefont{H.}~\bibnamefont{{Shan}}},
  \bibinfo{journal}{arXiv e-prints} \bibinfo{eid}{arXiv:2310.15053}
  (\bibinfo{year}{2023}), \eprint{2310.15053}.

\bibitem[{\citenamefont{{Qin} et~al.}(2024)\citenamefont{{Qin}, {Zhang}, {Yu},
  {Xu}, {Yao}, {Shi}, and {Shan}}}]{Qin2+}
\bibinfo{author}{\bibfnamefont{J.}~\bibnamefont{{Qin}}},
  \bibinfo{author}{\bibfnamefont{P.}~\bibnamefont{{Zhang}}},
  \bibinfo{author}{\bibfnamefont{Y.}~\bibnamefont{{Yu}}},
  \bibinfo{author}{\bibfnamefont{H.}~\bibnamefont{{Xu}}},
  \bibinfo{author}{\bibfnamefont{J.}~\bibnamefont{{Yao}}},
  \bibinfo{author}{\bibfnamefont{Y.}~\bibnamefont{{Shi}}}, \bibnamefont{and}
  \bibinfo{author}{\bibfnamefont{H.}~\bibnamefont{{Shan}}},
  \bibinfo{journal}{arXiv e-prints} \bibinfo{eid}{arXiv:2412.00829}
  (\bibinfo{year}{2024}), \eprint{2412.00829}.

\bibitem[{\citenamefont{{Bartelmann} and
  {Schneider}}(2001)}]{2001PhR...340..291B}
\bibinfo{author}{\bibfnamefont{M.}~\bibnamefont{{Bartelmann}}}
  \bibnamefont{and}
  \bibinfo{author}{\bibfnamefont{P.}~\bibnamefont{{Schneider}}},
  \bibinfo{journal}{\physrep} \textbf{\bibinfo{volume}{340}},
  \bibinfo{pages}{291} (\bibinfo{year}{2001}), \eprint{astro-ph/9912508}.

\bibitem[{\citenamefont{{Kilbinger}}(2015)}]{2015RPPh...78h6901K}
\bibinfo{author}{\bibfnamefont{M.}~\bibnamefont{{Kilbinger}}},
  \bibinfo{journal}{Reports on Progress in Physics}
  \textbf{\bibinfo{volume}{78}}, \bibinfo{eid}{086901} (\bibinfo{year}{2015}),
  \eprint{1411.0115}.

\bibitem[{\citenamefont{{Hamana} et~al.}(2020)\citenamefont{{Hamana},
  {Shirasaki}, {Miyazaki}, {Hikage}, {Oguri}, {More}, {Armstrong}, {Leauthaud},
  {Mandelbaum}, {Miyatake} et~al.}}]{Hamana2020}
\bibinfo{author}{\bibfnamefont{T.}~\bibnamefont{{Hamana}}},
  \bibinfo{author}{\bibfnamefont{M.}~\bibnamefont{{Shirasaki}}},
  \bibinfo{author}{\bibfnamefont{S.}~\bibnamefont{{Miyazaki}}},
  \bibinfo{author}{\bibfnamefont{C.}~\bibnamefont{{Hikage}}},
  \bibinfo{author}{\bibfnamefont{M.}~\bibnamefont{{Oguri}}},
  \bibinfo{author}{\bibfnamefont{S.}~\bibnamefont{{More}}},
  \bibinfo{author}{\bibfnamefont{R.}~\bibnamefont{{Armstrong}}},
  \bibinfo{author}{\bibfnamefont{A.}~\bibnamefont{{Leauthaud}}},
  \bibinfo{author}{\bibfnamefont{R.}~\bibnamefont{{Mandelbaum}}},
  \bibinfo{author}{\bibfnamefont{H.}~\bibnamefont{{Miyatake}}},
  \bibnamefont{et~al.}, \bibinfo{journal}{\pasj} \textbf{\bibinfo{volume}{72}},
  \bibinfo{eid}{16} (\bibinfo{year}{2020}), \eprint{1906.06041}.

\bibitem[{\citenamefont{{Asgari} et~al.}(2021)\citenamefont{{Asgari}, {Lin},
  {Joachimi}, {Giblin}, {Heymans}, {Hildebrandt}, {Kannawadi}, {St{\"o}lzner},
  {Tr{\"o}ster}, {van den Busch} et~al.}}]{2021A&A...645A.104A}
\bibinfo{author}{\bibfnamefont{M.}~\bibnamefont{{Asgari}}},
  \bibinfo{author}{\bibfnamefont{C.-A.} \bibnamefont{{Lin}}},
  \bibinfo{author}{\bibfnamefont{B.}~\bibnamefont{{Joachimi}}},
  \bibinfo{author}{\bibfnamefont{B.}~\bibnamefont{{Giblin}}},
  \bibinfo{author}{\bibfnamefont{C.}~\bibnamefont{{Heymans}}},
  \bibinfo{author}{\bibfnamefont{H.}~\bibnamefont{{Hildebrandt}}},
  \bibinfo{author}{\bibfnamefont{A.}~\bibnamefont{{Kannawadi}}},
  \bibinfo{author}{\bibfnamefont{B.}~\bibnamefont{{St{\"o}lzner}}},
  \bibinfo{author}{\bibfnamefont{T.}~\bibnamefont{{Tr{\"o}ster}}},
  \bibinfo{author}{\bibfnamefont{J.~L.} \bibnamefont{{van den Busch}}},
  \bibnamefont{et~al.}, \bibinfo{journal}{\aap} \textbf{\bibinfo{volume}{645}},
  \bibinfo{eid}{A104} (\bibinfo{year}{2021}), \eprint{2007.15633}.

\bibitem[{\citenamefont{{Giblin} et~al.}(2021)\citenamefont{{Giblin},
  {Heymans}, {Asgari}, {Hildebrandt}, {Hoekstra}, {Joachimi}, {Kannawadi},
  {Kuijken}, {Lin}, {Miller} et~al.}}]{2021A&A...645A.105G}
\bibinfo{author}{\bibfnamefont{B.}~\bibnamefont{{Giblin}}},
  \bibinfo{author}{\bibfnamefont{C.}~\bibnamefont{{Heymans}}},
  \bibinfo{author}{\bibfnamefont{M.}~\bibnamefont{{Asgari}}},
  \bibinfo{author}{\bibfnamefont{H.}~\bibnamefont{{Hildebrandt}}},
  \bibinfo{author}{\bibfnamefont{H.}~\bibnamefont{{Hoekstra}}},
  \bibinfo{author}{\bibfnamefont{B.}~\bibnamefont{{Joachimi}}},
  \bibinfo{author}{\bibfnamefont{A.}~\bibnamefont{{Kannawadi}}},
  \bibinfo{author}{\bibfnamefont{K.}~\bibnamefont{{Kuijken}}},
  \bibinfo{author}{\bibfnamefont{C.-A.} \bibnamefont{{Lin}}},
  \bibinfo{author}{\bibfnamefont{L.}~\bibnamefont{{Miller}}},
  \bibnamefont{et~al.}, \bibinfo{journal}{\aap} \textbf{\bibinfo{volume}{645}},
  \bibinfo{eid}{A105} (\bibinfo{year}{2021}), \eprint{2007.01845}.

\bibitem[{\citenamefont{{Loureiro} et~al.}(2022)\citenamefont{{Loureiro},
  {Whittaker}, {Spurio Mancini}, {Joachimi}, {Cuceu}, {Asgari}, {St{\"o}lzner},
  {Tr{\"o}ster}, {Wright}, {Bilicki} et~al.}}]{2022A&A...665A..56L}
\bibinfo{author}{\bibfnamefont{A.}~\bibnamefont{{Loureiro}}},
  \bibinfo{author}{\bibfnamefont{L.}~\bibnamefont{{Whittaker}}},
  \bibinfo{author}{\bibfnamefont{A.}~\bibnamefont{{Spurio Mancini}}},
  \bibinfo{author}{\bibfnamefont{B.}~\bibnamefont{{Joachimi}}},
  \bibinfo{author}{\bibfnamefont{A.}~\bibnamefont{{Cuceu}}},
  \bibinfo{author}{\bibfnamefont{M.}~\bibnamefont{{Asgari}}},
  \bibinfo{author}{\bibfnamefont{B.}~\bibnamefont{{St{\"o}lzner}}},
  \bibinfo{author}{\bibfnamefont{T.}~\bibnamefont{{Tr{\"o}ster}}},
  \bibinfo{author}{\bibfnamefont{A.~H.} \bibnamefont{{Wright}}},
  \bibinfo{author}{\bibfnamefont{M.}~\bibnamefont{{Bilicki}}},
  \bibnamefont{et~al.}, \bibinfo{journal}{\aap} \textbf{\bibinfo{volume}{665}},
  \bibinfo{eid}{A56} (\bibinfo{year}{2022}), \eprint{2110.06947}.

\bibitem[{\citenamefont{Amon et~al.}(2022)\citenamefont{Amon, Gruen, Troxel,
  MacCrann, Dodelson, Choi, Doux, Secco, Samuroff, Krause
  et~al.}}]{PhysRevD.105.023514}
\bibinfo{author}{\bibfnamefont{A.}~\bibnamefont{Amon}},
  \bibinfo{author}{\bibfnamefont{D.}~\bibnamefont{Gruen}},
  \bibinfo{author}{\bibfnamefont{M.~A.} \bibnamefont{Troxel}},
  \bibinfo{author}{\bibfnamefont{N.}~\bibnamefont{MacCrann}},
  \bibinfo{author}{\bibfnamefont{S.}~\bibnamefont{Dodelson}},
  \bibinfo{author}{\bibfnamefont{A.}~\bibnamefont{Choi}},
  \bibinfo{author}{\bibfnamefont{C.}~\bibnamefont{Doux}},
  \bibinfo{author}{\bibfnamefont{L.~F.} \bibnamefont{Secco}},
  \bibinfo{author}{\bibfnamefont{S.}~\bibnamefont{Samuroff}},
  \bibinfo{author}{\bibfnamefont{E.}~\bibnamefont{Krause}},
  \bibnamefont{et~al.} (\bibinfo{collaboration}{DES Collaboration}),
  \bibinfo{journal}{Phys. Rev. D} \textbf{\bibinfo{volume}{105}},
  \bibinfo{pages}{023514} (\bibinfo{year}{2022}),
  \urlprefix\url{https://link.aps.org/doi/10.1103/PhysRevD.105.023514}.

\bibitem[{\citenamefont{Secco et~al.}(2022)\citenamefont{Secco, Samuroff,
  Krause, Jain, Blazek, Raveri, Campos, Amon, Chen, Doux
  et~al.}}]{PhysRevD.105.023515}
\bibinfo{author}{\bibfnamefont{L.~F.} \bibnamefont{Secco}},
  \bibinfo{author}{\bibfnamefont{S.}~\bibnamefont{Samuroff}},
  \bibinfo{author}{\bibfnamefont{E.}~\bibnamefont{Krause}},
  \bibinfo{author}{\bibfnamefont{B.}~\bibnamefont{Jain}},
  \bibinfo{author}{\bibfnamefont{J.}~\bibnamefont{Blazek}},
  \bibinfo{author}{\bibfnamefont{M.}~\bibnamefont{Raveri}},
  \bibinfo{author}{\bibfnamefont{A.}~\bibnamefont{Campos}},
  \bibinfo{author}{\bibfnamefont{A.}~\bibnamefont{Amon}},
  \bibinfo{author}{\bibfnamefont{A.}~\bibnamefont{Chen}},
  \bibinfo{author}{\bibfnamefont{C.}~\bibnamefont{Doux}}, \bibnamefont{et~al.}
  (\bibinfo{collaboration}{DES Collaboration}), \bibinfo{journal}{Phys. Rev. D}
  \textbf{\bibinfo{volume}{105}}, \bibinfo{pages}{023515}
  (\bibinfo{year}{2022}),
  \urlprefix\url{https://link.aps.org/doi/10.1103/PhysRevD.105.023515}.

\bibitem[{\citenamefont{{Li} et~al.}(2023)\citenamefont{{Li}, {Zhang},
  {Sugiyama}, {Dalal}, {Rau}, {Mandelbaum}, {Takada}, {More}, {Strauss},
  {Miyatake} et~al.}}]{2023arXiv230400702L}
\bibinfo{author}{\bibfnamefont{X.}~\bibnamefont{{Li}}},
  \bibinfo{author}{\bibfnamefont{T.}~\bibnamefont{{Zhang}}},
  \bibinfo{author}{\bibfnamefont{S.}~\bibnamefont{{Sugiyama}}},
  \bibinfo{author}{\bibfnamefont{R.}~\bibnamefont{{Dalal}}},
  \bibinfo{author}{\bibfnamefont{M.~M.} \bibnamefont{{Rau}}},
  \bibinfo{author}{\bibfnamefont{R.}~\bibnamefont{{Mandelbaum}}},
  \bibinfo{author}{\bibfnamefont{M.}~\bibnamefont{{Takada}}},
  \bibinfo{author}{\bibfnamefont{S.}~\bibnamefont{{More}}},
  \bibinfo{author}{\bibfnamefont{M.~A.} \bibnamefont{{Strauss}}},
  \bibinfo{author}{\bibfnamefont{H.}~\bibnamefont{{Miyatake}}},
  \bibnamefont{et~al.}, \bibinfo{journal}{arXiv e-prints}
  \bibinfo{eid}{arXiv:2304.00702} (\bibinfo{year}{2023}), \eprint{2304.00702}.

\bibitem[{\citenamefont{{Dark Energy Survey Collaboration}
  et~al.}(2016)\citenamefont{{Dark Energy Survey Collaboration}, {Abbott},
  {Abdalla}, {Aleksi{\'c}}, {Allam}, {Amara}, {Bacon}, {Balbinot}, {Banerji},
  {Bechtol} et~al.}}]{DES2016}
\bibinfo{author}{\bibnamefont{{Dark Energy Survey Collaboration}}},
  \bibinfo{author}{\bibfnamefont{T.}~\bibnamefont{{Abbott}}},
  \bibinfo{author}{\bibfnamefont{F.~B.} \bibnamefont{{Abdalla}}},
  \bibinfo{author}{\bibfnamefont{J.}~\bibnamefont{{Aleksi{\'c}}}},
  \bibinfo{author}{\bibfnamefont{S.}~\bibnamefont{{Allam}}},
  \bibinfo{author}{\bibfnamefont{A.}~\bibnamefont{{Amara}}},
  \bibinfo{author}{\bibfnamefont{D.}~\bibnamefont{{Bacon}}},
  \bibinfo{author}{\bibfnamefont{E.}~\bibnamefont{{Balbinot}}},
  \bibinfo{author}{\bibfnamefont{M.}~\bibnamefont{{Banerji}}},
  \bibinfo{author}{\bibfnamefont{K.}~\bibnamefont{{Bechtol}}},
  \bibnamefont{et~al.}, \bibinfo{journal}{\mnras}
  \textbf{\bibinfo{volume}{460}}, \bibinfo{pages}{1270} (\bibinfo{year}{2016}),
  \eprint{1601.00329}.

\bibitem[{\citenamefont{{de Jong} et~al.}(2013)\citenamefont{{de Jong},
  {Verdoes Kleijn}, {Kuijken}, and {Valentijn}}}]{KiDS2013}
\bibinfo{author}{\bibfnamefont{J.~T.~A.} \bibnamefont{{de Jong}}},
  \bibinfo{author}{\bibfnamefont{G.~A.} \bibnamefont{{Verdoes Kleijn}}},
  \bibinfo{author}{\bibfnamefont{K.~H.} \bibnamefont{{Kuijken}}},
  \bibnamefont{and} \bibinfo{author}{\bibfnamefont{E.~A.}
  \bibnamefont{{Valentijn}}}, \bibinfo{journal}{Experimental Astronomy}
  \textbf{\bibinfo{volume}{35}}, \bibinfo{pages}{25} (\bibinfo{year}{2013}),
  \eprint{1206.1254}.

\bibitem[{\citenamefont{{Aihara} et~al.}(2018)\citenamefont{{Aihara},
  {Arimoto}, {Armstrong}, {Arnouts}, {Bahcall}, {Bickerton}, {Bosch}, {Bundy},
  {Capak}, {Chan} et~al.}}]{HSC2018}
\bibinfo{author}{\bibfnamefont{H.}~\bibnamefont{{Aihara}}},
  \bibinfo{author}{\bibfnamefont{N.}~\bibnamefont{{Arimoto}}},
  \bibinfo{author}{\bibfnamefont{R.}~\bibnamefont{{Armstrong}}},
  \bibinfo{author}{\bibfnamefont{S.}~\bibnamefont{{Arnouts}}},
  \bibinfo{author}{\bibfnamefont{N.~A.} \bibnamefont{{Bahcall}}},
  \bibinfo{author}{\bibfnamefont{S.}~\bibnamefont{{Bickerton}}},
  \bibinfo{author}{\bibfnamefont{J.}~\bibnamefont{{Bosch}}},
  \bibinfo{author}{\bibfnamefont{K.}~\bibnamefont{{Bundy}}},
  \bibinfo{author}{\bibfnamefont{P.~L.} \bibnamefont{{Capak}}},
  \bibinfo{author}{\bibfnamefont{J.~H.~H.} \bibnamefont{{Chan}}},
  \bibnamefont{et~al.}, \bibinfo{journal}{\pasj} \textbf{\bibinfo{volume}{70}},
  \bibinfo{eid}{S4} (\bibinfo{year}{2018}), \eprint{1704.05858}.

\bibitem[{\citenamefont{{LSST Science Collaboration}
  et~al.}(2009)\citenamefont{{LSST Science Collaboration}, {Abell}, {Allison},
  {Anderson}, {Andrew}, {Angel}, {Armus}, {Arnett}, {Asztalos}, {Axelrod}
  et~al.}}]{2009arXiv0912.0201L}
\bibinfo{author}{\bibnamefont{{LSST Science Collaboration}}},
  \bibinfo{author}{\bibfnamefont{P.~A.} \bibnamefont{{Abell}}},
  \bibinfo{author}{\bibfnamefont{J.}~\bibnamefont{{Allison}}},
  \bibinfo{author}{\bibfnamefont{S.~F.} \bibnamefont{{Anderson}}},
  \bibinfo{author}{\bibfnamefont{J.~R.} \bibnamefont{{Andrew}}},
  \bibinfo{author}{\bibfnamefont{J.~R.~P.} \bibnamefont{{Angel}}},
  \bibinfo{author}{\bibfnamefont{L.}~\bibnamefont{{Armus}}},
  \bibinfo{author}{\bibfnamefont{D.}~\bibnamefont{{Arnett}}},
  \bibinfo{author}{\bibfnamefont{S.~J.} \bibnamefont{{Asztalos}}},
  \bibinfo{author}{\bibfnamefont{T.~S.} \bibnamefont{{Axelrod}}},
  \bibnamefont{et~al.}, \bibinfo{journal}{arXiv e-prints}
  \bibinfo{eid}{arXiv:0912.0201} (\bibinfo{year}{2009}), \eprint{0912.0201}.

\bibitem[{\citenamefont{{Laureijs} et~al.}(2011)\citenamefont{{Laureijs},
  {Amiaux}, {Arduini}, {Augu{\`e}res}, {Brinchmann}, {Cole}, {Cropper},
  {Dabin}, {Duvet}, {Ealet} et~al.}}]{2011arXiv1110.3193L}
\bibinfo{author}{\bibfnamefont{R.}~\bibnamefont{{Laureijs}}},
  \bibinfo{author}{\bibfnamefont{J.}~\bibnamefont{{Amiaux}}},
  \bibinfo{author}{\bibfnamefont{S.}~\bibnamefont{{Arduini}}},
  \bibinfo{author}{\bibfnamefont{J.~L.} \bibnamefont{{Augu{\`e}res}}},
  \bibinfo{author}{\bibfnamefont{J.}~\bibnamefont{{Brinchmann}}},
  \bibinfo{author}{\bibfnamefont{R.}~\bibnamefont{{Cole}}},
  \bibinfo{author}{\bibfnamefont{M.}~\bibnamefont{{Cropper}}},
  \bibinfo{author}{\bibfnamefont{C.}~\bibnamefont{{Dabin}}},
  \bibinfo{author}{\bibfnamefont{L.}~\bibnamefont{{Duvet}}},
  \bibinfo{author}{\bibfnamefont{A.}~\bibnamefont{{Ealet}}},
  \bibnamefont{et~al.}, \bibinfo{journal}{arXiv e-prints}
  \bibinfo{eid}{arXiv:1110.3193} (\bibinfo{year}{2011}), \eprint{1110.3193}.

\bibitem[{\citenamefont{{Gong} et~al.}(2019)\citenamefont{{Gong}, {Liu}, {Cao},
  {Chen}, {Fan}, {Li}, {Li}, {Li}, {Zhang}, and {Zhan}}}]{2019ApJ...883..203G}
\bibinfo{author}{\bibfnamefont{Y.}~\bibnamefont{{Gong}}},
  \bibinfo{author}{\bibfnamefont{X.}~\bibnamefont{{Liu}}},
  \bibinfo{author}{\bibfnamefont{Y.}~\bibnamefont{{Cao}}},
  \bibinfo{author}{\bibfnamefont{X.}~\bibnamefont{{Chen}}},
  \bibinfo{author}{\bibfnamefont{Z.}~\bibnamefont{{Fan}}},
  \bibinfo{author}{\bibfnamefont{R.}~\bibnamefont{{Li}}},
  \bibinfo{author}{\bibfnamefont{X.-D.} \bibnamefont{{Li}}},
  \bibinfo{author}{\bibfnamefont{Z.}~\bibnamefont{{Li}}},
  \bibinfo{author}{\bibfnamefont{X.}~\bibnamefont{{Zhang}}}, \bibnamefont{and}
  \bibinfo{author}{\bibfnamefont{H.}~\bibnamefont{{Zhan}}},
  \bibinfo{journal}{\apj} \textbf{\bibinfo{volume}{883}}, \bibinfo{eid}{203}
  (\bibinfo{year}{2019}), \eprint{1901.04634}.

\bibitem[{\citenamefont{{Yao} et~al.}(2024)\citenamefont{{Yao}, {Shan}, {Li},
  {Xu}, {Fan}, {Liu}, {Zhang}, {Yu}, {Wei}, {Hu} et~al.}}]{2024MNRAS.527.5206Y}
\bibinfo{author}{\bibfnamefont{J.}~\bibnamefont{{Yao}}},
  \bibinfo{author}{\bibfnamefont{H.}~\bibnamefont{{Shan}}},
  \bibinfo{author}{\bibfnamefont{R.}~\bibnamefont{{Li}}},
  \bibinfo{author}{\bibfnamefont{Y.}~\bibnamefont{{Xu}}},
  \bibinfo{author}{\bibfnamefont{D.}~\bibnamefont{{Fan}}},
  \bibinfo{author}{\bibfnamefont{D.}~\bibnamefont{{Liu}}},
  \bibinfo{author}{\bibfnamefont{P.}~\bibnamefont{{Zhang}}},
  \bibinfo{author}{\bibfnamefont{Y.}~\bibnamefont{{Yu}}},
  \bibinfo{author}{\bibfnamefont{C.}~\bibnamefont{{Wei}}},
  \bibinfo{author}{\bibfnamefont{B.}~\bibnamefont{{Hu}}}, \bibnamefont{et~al.},
  \bibinfo{journal}{\mnras} \textbf{\bibinfo{volume}{527}},
  \bibinfo{pages}{5206} (\bibinfo{year}{2024}), \eprint{2304.04489}.

\bibitem[{\citenamefont{Bauer et~al.}(2014)\citenamefont{Bauer, Gaztañaga,
  Martí, and Miquel}}]{pub.1059915677}
\bibinfo{author}{\bibfnamefont{A.~H.} \bibnamefont{Bauer}},
  \bibinfo{author}{\bibfnamefont{E.}~\bibnamefont{Gaztañaga}},
  \bibinfo{author}{\bibfnamefont{P.}~\bibnamefont{Martí}}, \bibnamefont{and}
  \bibinfo{author}{\bibfnamefont{R.}~\bibnamefont{Miquel}},
  \bibinfo{journal}{Monthly Notices of the Royal Astronomical Society}
  \textbf{\bibinfo{volume}{440}}, \bibinfo{pages}{3701} (\bibinfo{year}{2014}),
  \bibinfo{note}{https://academic.oup.com/mnras/article-pdf/440/4/3701/3913172/stu530.pdf},
  \urlprefix\url{https://app.dimensions.ai/details/publication/pub.1059915677}.

\bibitem[{\citenamefont{{Bellagamba} et~al.}(2019)\citenamefont{{Bellagamba},
  {Sereno}, {Roncarelli}, {Maturi}, {Radovich}, {Bardelli}, {Puddu},
  {Moscardini}, {Getman}, {Hildebrandt} et~al.}}]{2019MNRAS.484.1598B}
\bibinfo{author}{\bibfnamefont{F.}~\bibnamefont{{Bellagamba}}},
  \bibinfo{author}{\bibfnamefont{M.}~\bibnamefont{{Sereno}}},
  \bibinfo{author}{\bibfnamefont{M.}~\bibnamefont{{Roncarelli}}},
  \bibinfo{author}{\bibfnamefont{M.}~\bibnamefont{{Maturi}}},
  \bibinfo{author}{\bibfnamefont{M.}~\bibnamefont{{Radovich}}},
  \bibinfo{author}{\bibfnamefont{S.}~\bibnamefont{{Bardelli}}},
  \bibinfo{author}{\bibfnamefont{E.}~\bibnamefont{{Puddu}}},
  \bibinfo{author}{\bibfnamefont{L.}~\bibnamefont{{Moscardini}}},
  \bibinfo{author}{\bibfnamefont{F.}~\bibnamefont{{Getman}}},
  \bibinfo{author}{\bibfnamefont{H.}~\bibnamefont{{Hildebrandt}}},
  \bibnamefont{et~al.}, \bibinfo{journal}{\mnras}
  \textbf{\bibinfo{volume}{484}}, \bibinfo{pages}{1598} (\bibinfo{year}{2019}),
  \eprint{1810.02827}.

\bibitem[{\citenamefont{{Chiu} et~al.}(2016)\citenamefont{{Chiu}, {Dietrich},
  {Mohr}, {Applegate}, {Benson}, {Bleem}, {Bayliss}, {Bocquet}, {Carlstrom},
  {Capasso} et~al.}}]{2016MNRAS.457.3050C}
\bibinfo{author}{\bibfnamefont{I.}~\bibnamefont{{Chiu}}},
  \bibinfo{author}{\bibfnamefont{J.~P.} \bibnamefont{{Dietrich}}},
  \bibinfo{author}{\bibfnamefont{J.}~\bibnamefont{{Mohr}}},
  \bibinfo{author}{\bibfnamefont{D.~E.} \bibnamefont{{Applegate}}},
  \bibinfo{author}{\bibfnamefont{B.~A.} \bibnamefont{{Benson}}},
  \bibinfo{author}{\bibfnamefont{L.~E.} \bibnamefont{{Bleem}}},
  \bibinfo{author}{\bibfnamefont{M.~B.} \bibnamefont{{Bayliss}}},
  \bibinfo{author}{\bibfnamefont{S.}~\bibnamefont{{Bocquet}}},
  \bibinfo{author}{\bibfnamefont{J.~E.} \bibnamefont{{Carlstrom}}},
  \bibinfo{author}{\bibfnamefont{R.}~\bibnamefont{{Capasso}}},
  \bibnamefont{et~al.}, \bibinfo{journal}{\mnras}
  \textbf{\bibinfo{volume}{457}}, \bibinfo{pages}{3050} (\bibinfo{year}{2016}),
  \eprint{1510.01745}.

\bibitem[{\citenamefont{{Chiu} et~al.}(2020)\citenamefont{{Chiu}, {Umetsu},
  {Murata}, {Medezinski}, and {Oguri}}}]{2020MNRAS.495..428C}
\bibinfo{author}{\bibfnamefont{I.~N.} \bibnamefont{{Chiu}}},
  \bibinfo{author}{\bibfnamefont{K.}~\bibnamefont{{Umetsu}}},
  \bibinfo{author}{\bibfnamefont{R.}~\bibnamefont{{Murata}}},
  \bibinfo{author}{\bibfnamefont{E.}~\bibnamefont{{Medezinski}}},
  \bibnamefont{and} \bibinfo{author}{\bibfnamefont{M.}~\bibnamefont{{Oguri}}},
  \bibinfo{journal}{\mnras} \textbf{\bibinfo{volume}{495}},
  \bibinfo{pages}{428} (\bibinfo{year}{2020}), \eprint{1909.02042}.

\bibitem[{\citenamefont{{Scranton} et~al.}(2005)\citenamefont{{Scranton},
  {M{\'e}nard}, {Richards}, {Nichol}, {Myers}, {Jain}, {Gray}, {Bartelmann},
  {Brunner}, {Connolly} et~al.}}]{2005ApJ...633..589S}
\bibinfo{author}{\bibfnamefont{R.}~\bibnamefont{{Scranton}}},
  \bibinfo{author}{\bibfnamefont{B.}~\bibnamefont{{M{\'e}nard}}},
  \bibinfo{author}{\bibfnamefont{G.~T.} \bibnamefont{{Richards}}},
  \bibinfo{author}{\bibfnamefont{R.~C.} \bibnamefont{{Nichol}}},
  \bibinfo{author}{\bibfnamefont{A.~D.} \bibnamefont{{Myers}}},
  \bibinfo{author}{\bibfnamefont{B.}~\bibnamefont{{Jain}}},
  \bibinfo{author}{\bibfnamefont{A.}~\bibnamefont{{Gray}}},
  \bibinfo{author}{\bibfnamefont{M.}~\bibnamefont{{Bartelmann}}},
  \bibinfo{author}{\bibfnamefont{R.~J.} \bibnamefont{{Brunner}}},
  \bibinfo{author}{\bibfnamefont{A.~J.} \bibnamefont{{Connolly}}},
  \bibnamefont{et~al.}, \bibinfo{journal}{\apj} \textbf{\bibinfo{volume}{633}},
  \bibinfo{pages}{589} (\bibinfo{year}{2005}), \eprint{astro-ph/0504510}.

\bibitem[{\citenamefont{{Bauer} et~al.}(2012)\citenamefont{{Bauer}, {Baltay},
  {Ellman}, {Jerke}, {Rabinowitz}, and {Scalzo}}}]{2012ApJ...749...56B}
\bibinfo{author}{\bibfnamefont{A.~H.} \bibnamefont{{Bauer}}},
  \bibinfo{author}{\bibfnamefont{C.}~\bibnamefont{{Baltay}}},
  \bibinfo{author}{\bibfnamefont{N.}~\bibnamefont{{Ellman}}},
  \bibinfo{author}{\bibfnamefont{J.}~\bibnamefont{{Jerke}}},
  \bibinfo{author}{\bibfnamefont{D.}~\bibnamefont{{Rabinowitz}}},
  \bibnamefont{and} \bibinfo{author}{\bibfnamefont{R.}~\bibnamefont{{Scalzo}}},
  \bibinfo{journal}{\apj} \textbf{\bibinfo{volume}{749}}, \bibinfo{eid}{56}
  (\bibinfo{year}{2012}), \eprint{1202.1371}.

\bibitem[{\citenamefont{{Morrison} et~al.}(2012)\citenamefont{{Morrison},
  {Scranton}, {M{\'e}nard}, {Schmidt}, {Tyson}, {Ryan}, {Choi}, and
  {Wittman}}}]{2012MNRAS.426.2489M}
\bibinfo{author}{\bibfnamefont{C.~B.} \bibnamefont{{Morrison}}},
  \bibinfo{author}{\bibfnamefont{R.}~\bibnamefont{{Scranton}}},
  \bibinfo{author}{\bibfnamefont{B.}~\bibnamefont{{M{\'e}nard}}},
  \bibinfo{author}{\bibfnamefont{S.~J.} \bibnamefont{{Schmidt}}},
  \bibinfo{author}{\bibfnamefont{J.~A.} \bibnamefont{{Tyson}}},
  \bibinfo{author}{\bibfnamefont{R.}~\bibnamefont{{Ryan}}},
  \bibinfo{author}{\bibfnamefont{A.}~\bibnamefont{{Choi}}}, \bibnamefont{and}
  \bibinfo{author}{\bibfnamefont{D.~M.} \bibnamefont{{Wittman}}},
  \bibinfo{journal}{\mnras} \textbf{\bibinfo{volume}{426}},
  \bibinfo{pages}{2489} (\bibinfo{year}{2012}), \eprint{1204.2830}.

\bibitem[{\citenamefont{{Tudorica} et~al.}(2017)\citenamefont{{Tudorica},
  {Hildebrandt}, {Tewes}, {Hoekstra}, {Morrison}, {Muzzin}, {Wilson}, {Yee},
  {Lidman}, {Hicks} et~al.}}]{2017A&A...608A.141T}
\bibinfo{author}{\bibfnamefont{A.}~\bibnamefont{{Tudorica}}},
  \bibinfo{author}{\bibfnamefont{H.}~\bibnamefont{{Hildebrandt}}},
  \bibinfo{author}{\bibfnamefont{M.}~\bibnamefont{{Tewes}}},
  \bibinfo{author}{\bibfnamefont{H.}~\bibnamefont{{Hoekstra}}},
  \bibinfo{author}{\bibfnamefont{C.~B.} \bibnamefont{{Morrison}}},
  \bibinfo{author}{\bibfnamefont{A.}~\bibnamefont{{Muzzin}}},
  \bibinfo{author}{\bibfnamefont{G.}~\bibnamefont{{Wilson}}},
  \bibinfo{author}{\bibfnamefont{H.~K.~C.} \bibnamefont{{Yee}}},
  \bibinfo{author}{\bibfnamefont{C.}~\bibnamefont{{Lidman}}},
  \bibinfo{author}{\bibfnamefont{A.}~\bibnamefont{{Hicks}}},
  \bibnamefont{et~al.}, \bibinfo{journal}{\aap} \textbf{\bibinfo{volume}{608}},
  \bibinfo{eid}{A141} (\bibinfo{year}{2017}), \eprint{1710.06431}.

\bibitem[{\citenamefont{{Bonavera} et~al.}(2021)\citenamefont{{Bonavera},
  {Cueli}, {Gonz{\'a}lez-Nuevo}, {Ronconi}, {Migliaccio}, {Lapi}, {Casas}, and
  {Crespo}}}]{2021A&A...656A..99B}
\bibinfo{author}{\bibfnamefont{L.}~\bibnamefont{{Bonavera}}},
  \bibinfo{author}{\bibfnamefont{M.~M.} \bibnamefont{{Cueli}}},
  \bibinfo{author}{\bibfnamefont{J.}~\bibnamefont{{Gonz{\'a}lez-Nuevo}}},
  \bibinfo{author}{\bibfnamefont{T.}~\bibnamefont{{Ronconi}}},
  \bibinfo{author}{\bibfnamefont{M.}~\bibnamefont{{Migliaccio}}},
  \bibinfo{author}{\bibfnamefont{A.}~\bibnamefont{{Lapi}}},
  \bibinfo{author}{\bibfnamefont{J.~M.} \bibnamefont{{Casas}}},
  \bibnamefont{and} \bibinfo{author}{\bibfnamefont{D.}~\bibnamefont{{Crespo}}},
  \bibinfo{journal}{\aap} \textbf{\bibinfo{volume}{656}}, \bibinfo{eid}{A99}
  (\bibinfo{year}{2021}), \eprint{2109.12413}.

\bibitem[{\citenamefont{{Crespo} et~al.}(2022)\citenamefont{{Crespo},
  {Gonz{\'a}lez-Nuevo}, {Bonavera}, {Cueli}, {Casas}, and
  {Goitia}}}]{Crespo_GonzalezNuevo_Bonavera_Cueli_Casas_Goitia_2022}
\bibinfo{author}{\bibfnamefont{D.}~\bibnamefont{{Crespo}}},
  \bibinfo{author}{\bibfnamefont{J.}~\bibnamefont{{Gonz{\'a}lez-Nuevo}}},
  \bibinfo{author}{\bibfnamefont{L.}~\bibnamefont{{Bonavera}}},
  \bibinfo{author}{\bibfnamefont{M.~M.} \bibnamefont{{Cueli}}},
  \bibinfo{author}{\bibfnamefont{J.~M.} \bibnamefont{{Casas}}},
  \bibnamefont{and} \bibinfo{author}{\bibfnamefont{E.}~\bibnamefont{{Goitia}}},
  \bibinfo{journal}{\aap} \textbf{\bibinfo{volume}{667}}, \bibinfo{eid}{A146}
  (\bibinfo{year}{2022}), \eprint{2210.17318}.

\bibitem[{\citenamefont{{Zhang} and {Pen}}(2005)}]{2005PhRvL..95x1302Z}
\bibinfo{author}{\bibfnamefont{P.}~\bibnamefont{{Zhang}}} \bibnamefont{and}
  \bibinfo{author}{\bibfnamefont{U.-L.} \bibnamefont{{Pen}}},
  \bibinfo{journal}{\prl} \textbf{\bibinfo{volume}{95}}, \bibinfo{eid}{241302}
  (\bibinfo{year}{2005}), \eprint{astro-ph/0506740}.

\bibitem[{\citenamefont{{Yang} and {Zhang}}(2011)}]{2011MNRAS.415.3485Y}
\bibinfo{author}{\bibfnamefont{X.}~\bibnamefont{{Yang}}} \bibnamefont{and}
  \bibinfo{author}{\bibfnamefont{P.}~\bibnamefont{{Zhang}}},
  \bibinfo{journal}{\mnras} \textbf{\bibinfo{volume}{415}},
  \bibinfo{pages}{3485} (\bibinfo{year}{2011}), \eprint{1105.2385}.

\bibitem[{\citenamefont{{Zhang} et~al.}(2019)\citenamefont{{Zhang}, {Zhang},
  and {Zhang}}}]{ABS}
\bibinfo{author}{\bibfnamefont{P.}~\bibnamefont{{Zhang}}},
  \bibinfo{author}{\bibfnamefont{J.}~\bibnamefont{{Zhang}}}, \bibnamefont{and}
  \bibinfo{author}{\bibfnamefont{L.}~\bibnamefont{{Zhang}}},
  \bibinfo{journal}{\mnras} \textbf{\bibinfo{volume}{484}},
  \bibinfo{pages}{1616} (\bibinfo{year}{2019}).

\bibitem[{\citenamefont{{Yang} et~al.}(2015)\citenamefont{{Yang}, {Zhang},
  {Zhang}, and {Yu}}}]{YangXJ15}
\bibinfo{author}{\bibfnamefont{X.}~\bibnamefont{{Yang}}},
  \bibinfo{author}{\bibfnamefont{P.}~\bibnamefont{{Zhang}}},
  \bibinfo{author}{\bibfnamefont{J.}~\bibnamefont{{Zhang}}}, \bibnamefont{and}
  \bibinfo{author}{\bibfnamefont{Y.}~\bibnamefont{{Yu}}},
  \bibinfo{journal}{\mnras} \textbf{\bibinfo{volume}{447}},
  \bibinfo{pages}{345} (\bibinfo{year}{2015}).

\bibitem[{\citenamefont{{Yang} et~al.}(2017)\citenamefont{{Yang}, {Zhang},
  {Yu}, and {Zhang}}}]{YangXJ17}
\bibinfo{author}{\bibfnamefont{X.}~\bibnamefont{{Yang}}},
  \bibinfo{author}{\bibfnamefont{J.}~\bibnamefont{{Zhang}}},
  \bibinfo{author}{\bibfnamefont{Y.}~\bibnamefont{{Yu}}}, \bibnamefont{and}
  \bibinfo{author}{\bibfnamefont{P.}~\bibnamefont{{Zhang}}},
  \bibinfo{journal}{\apj} \textbf{\bibinfo{volume}{845}}, \bibinfo{eid}{174}
  (\bibinfo{year}{2017}), \eprint{1703.01575}.

\bibitem[{\citenamefont{{Zhang} et~al.}(2018)\citenamefont{{Zhang}, {Yang},
  {Zhang}, and {Yu}}}]{Zhang18}
\bibinfo{author}{\bibfnamefont{P.}~\bibnamefont{{Zhang}}},
  \bibinfo{author}{\bibfnamefont{X.}~\bibnamefont{{Yang}}},
  \bibinfo{author}{\bibfnamefont{J.}~\bibnamefont{{Zhang}}}, \bibnamefont{and}
  \bibinfo{author}{\bibfnamefont{Y.}~\bibnamefont{{Yu}}},
  \bibinfo{journal}{\apj} \textbf{\bibinfo{volume}{864}}, \bibinfo{eid}{10}
  (\bibinfo{year}{2018}), \eprint{1807.00443}.

\bibitem[{\citenamefont{{Hou} et~al.}(2021)\citenamefont{{Hou}, {Yu}, and
  {Zhang}}}]{2021RAA....21..247H}
\bibinfo{author}{\bibfnamefont{S.-T.} \bibnamefont{{Hou}}},
  \bibinfo{author}{\bibfnamefont{Y.}~\bibnamefont{{Yu}}}, \bibnamefont{and}
  \bibinfo{author}{\bibfnamefont{P.-J.} \bibnamefont{{Zhang}}},
  \bibinfo{journal}{Research in Astronomy and Astrophysics}
  \textbf{\bibinfo{volume}{21}}, \bibinfo{eid}{247} (\bibinfo{year}{2021}),
  \eprint{2106.09970}.

\bibitem[{\citenamefont{{Ma} et~al.}(2024)\citenamefont{{Ma}, {Zhang}, {Yu},
  and {Qin}}}]{2024MNRAS.527.7547M}
\bibinfo{author}{\bibfnamefont{R.}~\bibnamefont{{Ma}}},
  \bibinfo{author}{\bibfnamefont{P.}~\bibnamefont{{Zhang}}},
  \bibinfo{author}{\bibfnamefont{Y.}~\bibnamefont{{Yu}}}, \bibnamefont{and}
  \bibinfo{author}{\bibfnamefont{J.}~\bibnamefont{{Qin}}},
  \bibinfo{journal}{\mnras} \textbf{\bibinfo{volume}{527}},
  \bibinfo{pages}{7547} (\bibinfo{year}{2024}), \eprint{2306.15177}.

\bibitem[{\citenamefont{{Bonoli} and {Pen}}(2009)}]{Bonoli09}
\bibinfo{author}{\bibfnamefont{S.}~\bibnamefont{{Bonoli}}} \bibnamefont{and}
  \bibinfo{author}{\bibfnamefont{U.~L.} \bibnamefont{{Pen}}},
  \bibinfo{journal}{\mnras} \textbf{\bibinfo{volume}{396}},
  \bibinfo{pages}{1610} (\bibinfo{year}{2009}), \eprint{0810.0273}.

\bibitem[{\citenamefont{{Hamaus} et~al.}(2010)\citenamefont{{Hamaus}, {Seljak},
  {Desjacques}, {Smith}, and {Baldauf}}}]{Hamaus10}
\bibinfo{author}{\bibfnamefont{N.}~\bibnamefont{{Hamaus}}},
  \bibinfo{author}{\bibfnamefont{U.}~\bibnamefont{{Seljak}}},
  \bibinfo{author}{\bibfnamefont{V.}~\bibnamefont{{Desjacques}}},
  \bibinfo{author}{\bibfnamefont{R.~E.} \bibnamefont{{Smith}}},
  \bibnamefont{and}
  \bibinfo{author}{\bibfnamefont{T.}~\bibnamefont{{Baldauf}}},
  \bibinfo{journal}{\prd} \textbf{\bibinfo{volume}{82}}, \bibinfo{eid}{043515}
  (\bibinfo{year}{2010}), \eprint{1004.5377}.

\bibitem[{\citenamefont{{Baldauf} et~al.}(2010)\citenamefont{{Baldauf},
  {Smith}, {Seljak}, and {Mandelbaum}}}]{Baldauf10}
\bibinfo{author}{\bibfnamefont{T.}~\bibnamefont{{Baldauf}}},
  \bibinfo{author}{\bibfnamefont{R.~E.} \bibnamefont{{Smith}}},
  \bibinfo{author}{\bibfnamefont{U.}~\bibnamefont{{Seljak}}}, \bibnamefont{and}
  \bibinfo{author}{\bibfnamefont{R.}~\bibnamefont{{Mandelbaum}}},
  \bibinfo{journal}{\prd} \textbf{\bibinfo{volume}{81}}, \bibinfo{eid}{063531}
  (\bibinfo{year}{2010}), \eprint{0911.4973}.

\bibitem[{\citenamefont{von Wietersheim Kramsta
  et~al.}(2021)\citenamefont{von Wietersheim Kramsta, Joachimi,
  van den Busch, Heymans, Hildebrandt, Asgari, Tr’oster, Unruh, and
  Wright}}]{Wietersheim-Kramsta_Joachimi_van}
\bibinfo{author}{\bibfnamefont{M.}~\bibnamefont{von Wietersheim Kramsta}},
  \bibinfo{author}{\bibfnamefont{B.}~\bibnamefont{Joachimi}},
  \bibinfo{author}{\bibfnamefont{J.~L.} \bibnamefont{van den Busch}},
  \bibinfo{author}{\bibfnamefont{C.}~\bibnamefont{Heymans}},
  \bibinfo{author}{\bibfnamefont{H.}~\bibnamefont{Hildebrandt}},
  \bibinfo{author}{\bibfnamefont{M.}~\bibnamefont{Asgari}},
  \bibinfo{author}{\bibfnamefont{T.}~\bibnamefont{Tr’oster}},
  \bibinfo{author}{\bibfnamefont{S.}~\bibnamefont{Unruh}}, \bibnamefont{and}
  \bibinfo{author}{\bibfnamefont{A.~H.} \bibnamefont{Wright}},
  \bibinfo{journal}{Monthly Notices of the Royal Astronomical Society}
  \textbf{\bibinfo{volume}{504}}, \bibinfo{pages}{1452–1465}
  (\bibinfo{year}{2021}),
  \urlprefix\url{http://dx.doi.org/10.1093/mnras/stab1000}.

\bibitem[{\citenamefont{Elvin-Poole et~al.}(2023)\citenamefont{Elvin-Poole,
  MacCrann, Everett, Prat, Rykoff, De Vicente, Yanny, Herner, Ferté,
  Valentino et~al.}}]{JElvinPoole2022DarkES}
\bibinfo{author}{\bibfnamefont{J.}~\bibnamefont{Elvin-Poole}},
  \bibinfo{author}{\bibfnamefont{N.}~\bibnamefont{MacCrann}},
  \bibinfo{author}{\bibfnamefont{S.}~\bibnamefont{Everett}},
  \bibinfo{author}{\bibfnamefont{J.}~\bibnamefont{Prat}},
  \bibinfo{author}{\bibfnamefont{E.~S.} \bibnamefont{Rykoff}},
  \bibinfo{author}{\bibfnamefont{J.}~\bibnamefont{De Vicente}},
  \bibinfo{author}{\bibfnamefont{B.}~\bibnamefont{Yanny}},
  \bibinfo{author}{\bibfnamefont{K.}~\bibnamefont{Herner}},
  \bibinfo{author}{\bibfnamefont{A.}~\bibnamefont{Ferté}},
  \bibinfo{author}{\bibfnamefont{E.~D.} \bibnamefont{Valentino}},
  \bibnamefont{et~al.}, \bibinfo{journal}{Monthly Notices of the Royal
  Astronomical Society} \textbf{\bibinfo{volume}{523}}, \bibinfo{pages}{3649}
  (\bibinfo{year}{2023}), ISSN \bibinfo{issn}{0035-8711},
  \eprint{https://academic.oup.com/mnras/article-pdf/523/3/3649/50596748/stad1594.pdf},
  \urlprefix\url{https://doi.org/10.1093/mnras/stad1594}.

\bibitem[{\citenamefont{{Wenzl} et~al.}(2024)\citenamefont{{Wenzl}, {Chen}, and
  {Bean}}}]{2024MNRAS.527.1760W}
\bibinfo{author}{\bibfnamefont{L.}~\bibnamefont{{Wenzl}}},
  \bibinfo{author}{\bibfnamefont{S.-F.} \bibnamefont{{Chen}}},
  \bibnamefont{and} \bibinfo{author}{\bibfnamefont{R.}~\bibnamefont{{Bean}}},
  \bibinfo{journal}{\mnras} \textbf{\bibinfo{volume}{527}},
  \bibinfo{pages}{1760} (\bibinfo{year}{2024}), \eprint{2308.05892}.

\bibitem[{\citenamefont{Zhou et~al.}(2021)\citenamefont{Zhou, Newman, Mao,
  Meisner, Moustakas, Myers, Prakash, Zentner, Brooks, Duan
  et~al.}}]{zhou2021clustering}
\bibinfo{author}{\bibfnamefont{R.}~\bibnamefont{Zhou}},
  \bibinfo{author}{\bibfnamefont{J.~A.} \bibnamefont{Newman}},
  \bibinfo{author}{\bibfnamefont{Y.-Y.} \bibnamefont{Mao}},
  \bibinfo{author}{\bibfnamefont{A.}~\bibnamefont{Meisner}},
  \bibinfo{author}{\bibfnamefont{J.}~\bibnamefont{Moustakas}},
  \bibinfo{author}{\bibfnamefont{A.~D.} \bibnamefont{Myers}},
  \bibinfo{author}{\bibfnamefont{A.}~\bibnamefont{Prakash}},
  \bibinfo{author}{\bibfnamefont{A.~R.} \bibnamefont{Zentner}},
  \bibinfo{author}{\bibfnamefont{D.}~\bibnamefont{Brooks}},
  \bibinfo{author}{\bibfnamefont{Y.}~\bibnamefont{Duan}}, \bibnamefont{et~al.},
  \bibinfo{journal}{Monthly Notices of the Royal Astronomical Society}
  \textbf{\bibinfo{volume}{501}}, \bibinfo{pages}{3309} (\bibinfo{year}{2021}).

\bibitem[{\citenamefont{{Joachimi} and {Bridle}}(2010)}]{2010A&A...523A...1J}
\bibinfo{author}{\bibfnamefont{B.}~\bibnamefont{{Joachimi}}} \bibnamefont{and}
  \bibinfo{author}{\bibfnamefont{S.~L.} \bibnamefont{{Bridle}}},
  \bibinfo{journal}{\aap} \textbf{\bibinfo{volume}{523}}, \bibinfo{eid}{A1}
  (\bibinfo{year}{2010}), \eprint{0911.2454}.

\bibitem[{\citenamefont{{Bernstein}}(2009)}]{2009ApJ...695..652B}
\bibinfo{author}{\bibfnamefont{G.~M.} \bibnamefont{{Bernstein}}},
  \bibinfo{journal}{\apj} \textbf{\bibinfo{volume}{695}}, \bibinfo{pages}{652}
  (\bibinfo{year}{2009}), \eprint{0808.3400}.

\bibitem[{\citenamefont{{Elvin-Poole} et~al.}(2023)\citenamefont{{Elvin-Poole},
  {MacCrann}, {Everett}, {Prat}, {Rykoff}, {De Vicente}, {Yanny}, {Herner},
  {Fert{\'e}}, {Di Valentino} et~al.}}]{2023MNRAS.523.3649E}
\bibinfo{author}{\bibfnamefont{J.}~\bibnamefont{{Elvin-Poole}}},
  \bibinfo{author}{\bibfnamefont{N.}~\bibnamefont{{MacCrann}}},
  \bibinfo{author}{\bibfnamefont{S.}~\bibnamefont{{Everett}}},
  \bibinfo{author}{\bibfnamefont{J.}~\bibnamefont{{Prat}}},
  \bibinfo{author}{\bibfnamefont{E.~S.} \bibnamefont{{Rykoff}}},
  \bibinfo{author}{\bibfnamefont{J.}~\bibnamefont{{De Vicente}}},
  \bibinfo{author}{\bibfnamefont{B.}~\bibnamefont{{Yanny}}},
  \bibinfo{author}{\bibfnamefont{K.}~\bibnamefont{{Herner}}},
  \bibinfo{author}{\bibfnamefont{A.}~\bibnamefont{{Fert{\'e}}}},
  \bibinfo{author}{\bibfnamefont{E.}~\bibnamefont{{Di Valentino}}},
  \bibnamefont{et~al.}, \bibinfo{journal}{\mnras}
  \textbf{\bibinfo{volume}{523}}, \bibinfo{pages}{3649} (\bibinfo{year}{2023}),
  \eprint{2209.09782}.

\bibitem[{\citenamefont{{Everett} et~al.}(2022)\citenamefont{{Everett},
  {Yanny}, {Kuropatkin}, {Huff}, {Zhang}, {Myles}, {Masegian}, {Elvin-Poole},
  {Allam}, {Bernstein} et~al.}}]{2022ApJS..258...15E}
\bibinfo{author}{\bibfnamefont{S.}~\bibnamefont{{Everett}}},
  \bibinfo{author}{\bibfnamefont{B.}~\bibnamefont{{Yanny}}},
  \bibinfo{author}{\bibfnamefont{N.}~\bibnamefont{{Kuropatkin}}},
  \bibinfo{author}{\bibfnamefont{E.~M.} \bibnamefont{{Huff}}},
  \bibinfo{author}{\bibfnamefont{Y.}~\bibnamefont{{Zhang}}},
  \bibinfo{author}{\bibfnamefont{J.}~\bibnamefont{{Myles}}},
  \bibinfo{author}{\bibfnamefont{A.}~\bibnamefont{{Masegian}}},
  \bibinfo{author}{\bibfnamefont{J.}~\bibnamefont{{Elvin-Poole}}},
  \bibinfo{author}{\bibfnamefont{S.}~\bibnamefont{{Allam}}},
  \bibinfo{author}{\bibfnamefont{G.~M.} \bibnamefont{{Bernstein}}},
  \bibnamefont{et~al.}, \bibinfo{journal}{\apjs}
  \textbf{\bibinfo{volume}{258}}, \bibinfo{eid}{15} (\bibinfo{year}{2022}),
  \eprint{2012.12825}.

\end{thebibliography}

%put the figures Clkk and Wiener, two plots in one row  

% \bsp	% typesetting comment
\label{lastpage}
\end{document}